\documentclass[acmtog, nonacm]{acmart}

\usepackage{booktabs} 

\usepackage[ruled]{algorithm2e} 

\SetAlFnt{\small}
\SetAlCapFnt{\small}
\SetAlCapNameFnt{\small}
\SetAlCapHSkip{0pt}

\makeatletter
\renewcommand{\@authorsaddresses}{\par\noindent
Authors' addresses: Yuze He, hyz22@mails.tsinghua.edu.cn, Tsinghua University, Beijing, China; Yanning Zhou, ynzhou0907@gmail.com, Tencent AIPD, Shenzhen, China; Wang Zhao, thuzhaowang@163.com, Tsinghua University, Beijing, China; Jingwen Ye, jingwenye@tencent.com, Tencent AIPD, Shenzhen, China; Yushi Bai, bys22@mails.tsinghua.edu.cn, Tsinghua University, Beijing, China; Kaiwen Xiao, scp173.cool@gmail.com, Tencent AIPD, Shenzhen, China; Yong-Jin Liu, liuyongjin@tsinghua.edu.cn, Tsinghua University, Beijing, China.
}
\makeatother

\newcommand{\xhdr}[1]{{\noindent\bfseries #1}.}
\newcommand{\ourmethod}{CHARM}

\acmJournal{TOG}

\begin{document}
\title{CHARM: Control-point-based 3D Anime Hairstyle Auto-Regressive Modeling}

\author{Yuze He}
\authornote{Work done during an internship at Tencent AIPD.}
\affiliation{
 \institution{Tsinghua University and Tencent AIPD}
 \city{Beijing}
 \country{China}
}
\email{hyz22@mails.tsinghua.edu.cn}

\author{Yanning Zhou}
\authornotemark[2]
\affiliation{
 \institution{Tencent AIPD}
 \city{Shenzhen}
 \country{China}}
\email{ynzhou0907@gmail.com}

\author{Wang Zhao}
\affiliation{
 \institution{Tsinghua University}
 \city{Beijing}
 \country{China}}
\email{thuzhaowang@163.com}

\author{Jingwen Ye}
\affiliation{
 \institution{Tencent AIPD}
 \city{Shenzhen}
 \country{China}}
\email{jingwenye@tencent.com}

\author{Yushi Bai}
\affiliation{
 \institution{Tsinghua University}
 \city{Beijing}
 \country{China}}
\email{bys22@mails.tsinghua.edu.cn}

\author{Kaiwen Xiao}
\affiliation{
 \institution{Tencent AIPD}
 \city{Shenzhen}
 \country{China}}
\email{scp173.cool@gmail.com}

\author{Yong-Jin Liu}
\authornote{Corresponding authors.}
\affiliation{
 \institution{Tsinghua University}
 \city{Beijing}
 \country{China}}
\email{liuyongjin@tsinghua.edu.cn}

\author{Zhongqian Sun}
\author{Wei Yang}
\affiliation{
 \institution{Tencent AIPD}
 \city{Shenzhen}
 \country{China}}

\begin{abstract}
We present \ourmethod, a novel parametric representation and generative framework for anime hairstyle modeling. While traditional hair modeling methods focus on realistic hair using strand-based or volumetric representations, anime hairstyle exhibits highly stylized, piecewise-structured geometry that challenges existing techniques. Existing works often rely on dense mesh modeling or hand-crafted spline curves, making them inefficient for editing and unsuitable for scalable learning. \ourmethod\ introduces a compact, invertible control-point-based parameterization, where a sequence of control points represents each hair card, and each point is encoded with only five geometric parameters. This efficient and accurate representation supports both artist-friendly design and learning-based generation.
Built upon this representation,  \ourmethod\ introduces an autoregressive generative framework that effectively generates anime hairstyles from input images or point clouds. By interpreting anime hairstyles as a sequential ``hair language'', our autoregressive transformer captures both local geometry and global hairstyle topology, resulting in high-fidelity anime hairstyle creation. To facilitate both training and evaluation of anime hairstyle generation, we construct AnimeHair, a large-scale dataset of 37K high-quality anime hairstyles with separated hair cards and processed mesh data. Extensive experiments demonstrate state-of-the-art performance of  \ourmethod\ in both reconstruction accuracy and generation quality, offering an expressive and scalable solution for anime hairstyle modeling. {\it Project page: https://hyzcluster.github.io/charm}
\end{abstract}

\begin{teaserfigure}
\setlength{\abovecaptionskip}{3pt}
    \centering
    \includegraphics[width=1\textwidth]{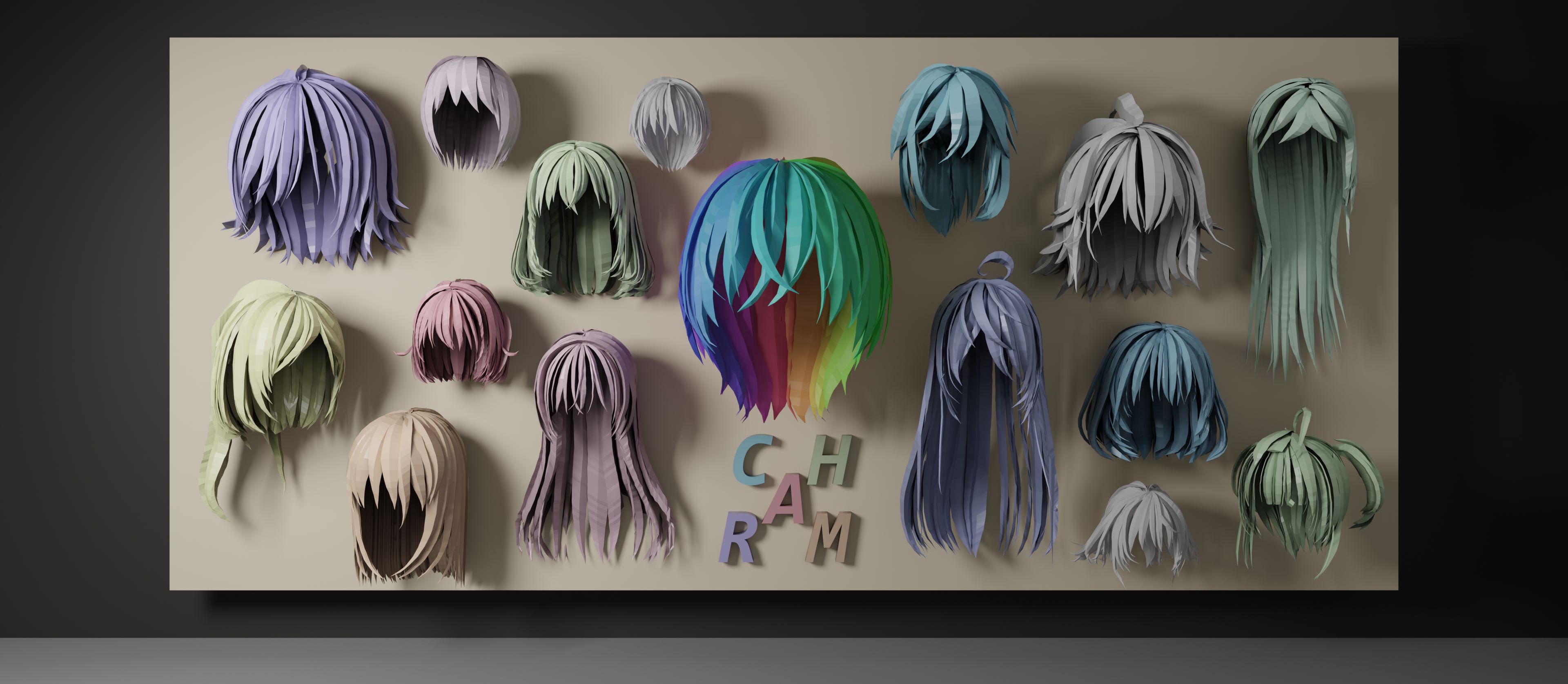}
    \caption{Based on a novel parametric representation, CHARM provides a generative framework for 3D anime hairstyle modeling.}
    \label{fig:teaser}
\end{teaserfigure}

\maketitle

\section{Introduction}
Hair modeling is a long-standing research challenge in computer graphics. It plays a vital role in creating expressive and immersive digital avatars for animation, filming, and games. Among various types of digital hairstyles, anime hairstyle has emerged as a distinct and popular category, characterized by its stylized, exaggerated shapes and vibrant aesthetics.

So far, modeling anime hairstyles remains underexplored. Traditional methods primarily focus on realistic hair, where a rich body of work has explored explicit (e.g. strands~\cite{koh2001simple,LiangH03, chen1999asystem,yang2000the}, hair mesh~\cite{yuksel2009hairmeshes}), implicit (e.g. implicit functions~\cite{rosu2022neuralstrands,sklyarova2023neural}) and hybrid (e.g. strand-guided 3D Gaussians~\cite{luo2024gaussianhair,zakharov2024human}) representations. In contrast, anime hairstyle is deliberately stylized with non-uniform strand thickness, variable hair density, non-fixed root positions and intra-hair structures, making it difficult to directly apply techniques developed for realistic hair.

Some existing character modeling methods~\cite{yan2024frankenstein, he2024stdgen} create anime hairstyles directly as a monolithic polygonal mesh. Despite generality and simplicity, the lack of hair structural separation significantly complicates further editing and animation. In practice, artists often rely on parametric primitives and curves, such as Bézier splines, to define anime hair shapes. 
While intuitive and flexible for manual design or interaction, they present challenges for automatic reconstruction and generation, since fitting Bézier curves or learning to create those is highly non-trivial. Thus, a compact, structured parametric representation for anime hairstyle is essential, yet still absent, for enabling scalable and learnable modeling pipelines.

To address this gap, we propose a novel parametric representation for anime hairstyles (Fig.~\ref{fig:teaser}). Our design is inspired by the modeling format adopted in the popular industry platform VRoid Hub~\cite{vroidHub}, which represents hair using repetitive geometric hair ``units'' (see Fig.~\ref{fig:unit}). Rather than directly modeling these units with full-resolution meshes—as done in VRoid, which is highly redundant and suboptimal for learning—we introduce a compact control-point-based parameterization. Each unit is defined by a control point with only five parameters indicating 3D position, width, and thickness. A sequence of such control points then constitutes a single hair card. By further developing an optimization system that supports conversion between our proposed parametric model and conventional 3D mesh formats, we achieve a fully invertible, editable, and lightweight parameterization for anime-style hair geometry. This parametric representation is efficient, accurate, and also intuitive to manipulate and well-suited for integration with data-driven generative models.

Building on this representation, we further introduce an autoregressive generative model for creating anime hairstyles from casual inputs, such as point clouds or images. Leveraging the sequential nature of our representation, we conceptualize the anime hairstyle as a \textit{hair language}, where each hair unit functions as a word, and each hair card as a sentence (see Fig.~\ref{fig:seq}). This language-like innovative representation for anime hairstyle naturally captures both intra-hair and inter-hair structural relationships, and facilitates the generation of varying hair card counts and lengths. An auto-regressive transformer network is designed to effectively learn to generate this geometric hair language, under the condition of point clouds or images.
Additionally, to train this model at scale, we construct AnimeHair, a large-scale dataset of 37K high-quality anime hairstyles derived from publicly available VRoid models. Each hairstyle is processed to isolate individual hair cards, and low-quality samples are filtered out. By training on this curated dataset, our model achieves high-quality and high-fidelity anime hairstyle generation, demonstrating the strong capability in both faithful reconstruction and creative hairstyle synthesis. Fig.~\ref{fig:teaser} shows some examples of our generated anime hairstyles.

Our contributions are summarized as follows: 1) We introduce a novel, efficient, and accurate parametric representation tailored for anime hairstyle modeling, which is not only easy to edit or interact by human users, but also structured and scalable for learning deep models to reconstruct or generate with, 2) We formulate the anime hairstyle generation as a sequential language-like task via the proposed parametric representation, and design an autoregressive transformer network that effectively learns to synthesize hairstyles from diverse inputs, 3) A carefully curated dataset of 37K anime hairstyles with separated hair cards and high-resolution meshes, providing a solid foundation for both training and evaluations of anime hairstyle generation, 4) Extensive experiments demonstrate that our system can generate high-fidelity anime hairs, establishing new state-of-the-art for anime hairstyle generation.
\section{Related Works}

\subsection{Hair Representation}

\textit{Realistic Hair Representation.}
Modeling realistic virtual hair has been widely studied in computer graphics. Due to the high geometric complexity of hair and the variety of hairstyles, most parametric methods are based on controlling collections of hair strands, where, for simplification, a strand is parameterized by a sequence of connected 3D points.
Various representations have been explored for hair strands, including parametric surfaces~\cite{koh2001simple,LiangH03,paul2006model}, generalized cylinders~\cite{chen1999asystem,yang2000the,xu2001vhair,kim2002iteractive}, and hair meshes~\cite{yuksel2009hairmeshes}. 
For characterizing individual hair strand geometry, researchers have proposed generalized helicoids~\cite{piuze2011generalized}, which represent both a single hair strand and its vicinity through three intuitive curvature parameters and an elevation angle.
Volumetric representation is also used for realistic-style hair modeling. ~\cite{saito2018hair} proposes to use volumetric orientation fields as an intermediate representation. 
However, the low resolution of voxels limits the fine detail reconstruction.
CT2Hair~\cite{shen2023CT2Hair} established a coarse-to-fine approach by creating density volumes of hair regions and estimating orientation fields to generate dense strands.
Recently, hair modeling methods incorporating neural rendering methods~\cite{wu2024monohair}.
GaussianHair~\cite{luo2024gaussianhair} and Zakharov et al.~\shortcite{zakharov2024human} design 3D Gaussians to represent the control points of the strands.
The works~\cite{sklyarova2023haar,rosu2022neuralstrands,sklyarova2023neural,he2024perm} utilize neural scalp textures in UV-space as the base hair representation for explicit 3D geometry and appearance of strands, employing various generative models for hairstyle synthesis.

\textit{Anime Hair Representation.}
Anime hairstyle modeling aims to support virtual avatar creation and computer-aided cartoon production by non-photorealistic rendering. 
Unlike realistic style modeling, where hair strand thickness is usually uniform, anime hairstyle is \textbf{deliberately stylized with varying strand thicknesses} to create the iconic clustered appearance essential to the anime aesthetic.
While most research has focused on photorealistic solutions, parametric modeling techniques specifically for anime hairstyles remain underexplored.
Some attempt to use implicit surfaces~\cite{sakai2013skeleton}, particle-based approaches~\cite{shin2006style}, and cluster polygons~\cite{mao2004sketch}. 
However, these approaches have shown limited effectiveness as they result in incomplete hairstyles and still require manual modeling by artists.
Another stream of works~\cite{mccan2009local,river2010cartoon,zhang2012excol,liu2013stereo,yeh20142} focused on 2.5D layered cartoon processing but lacked hair-specific design considerations or relied on 2D segmentation techniques that extract layers from reference views rather than generating complete 3D hairstyles.
Our representation for anime hairstyles captures diverse 3D shapes while preserving anime aesthetics, offering a compact and invertible structure that is well-suited for generative learning frameworks.

\subsection{Hair Modeling and Synthesis}
Hair representation presents unique challenges in computer graphics, requiring specialized techniques for both reconstruction from existing data and generation of new hairstyles. 
While general head reconstruction methods~\cite{gafni2021dynamic,giebenhain2023learning,hong2022headnerf,zhang2020h3dnet} incorporate hair, they typically produce only coarse outer shells with smooth geometry unsuitable for natural hair dynamics.

Specialized hair reconstruction approaches employ diverse techniques, from 3D orientation fields~\cite{nam2019strand,takimoto2024dr} to patch-based optimization~\cite{wu2024monohair} for high-fidelity exterior linemaps. 
Recent works have advanced along either designing more efficient representations~\cite{luo2024gaussianhair} or leveraging data-driven approaches~\cite{zakharov2024human,sklyarova2023neural} that incorporate priors to enhance reconstruction quality in challenging scenarios. However, these methods are primarily designed for realistic hair reconstruction and still face limitations when reconstructing complex occluded regions.

3D generation has witnessed remarkable progress, ranging from general object synthesis~\cite{zhang2024clay,petrov2024gem3d}, avatar creation~\cite{wang2024nova,he2024stdgen} to specialized realistic hair generation~\cite{zhou2023groomgen,he2024perm,sklyarova2023haar,long2025tangled}.
For realistic hair generation, NeuralStrands~\cite{rosu2022neuralstrands} and GroomGen~\cite{zhou2023groomgen} employ modulator-synthesizer architectures and GAN-based generators respectively.
Recently, HAAR~\cite{sklyarova2023haar} encodes hairstyle in UV feature space, and then leverage the latent diffusion model~\cite{podell2023sdxl} for strand generation.
TANGLED~\cite{long2025tangled} developed a multi-view lineart conditional diffusion approach followed by parametric braid inpainting. 
While it generates high-fidelity hair strands across diverse input styles, the resulting hair strands remain limited to realistic representations, which are unsuitable for anime character model requirements.

A key distinction exists between realistic and anime-style hair requirements. Anime hairstyle features deliberately stylized strands with non-uniform thickness, variable hair density, and irregular root positioning. These properties make fixed-resolution UV-space mapping unsuitable. StdGEN~\cite{he2024stdgen} attempts to generate decomposed anime hair meshes along with cloth and body, but it suffers from over-smoothing. 
Auto-regressive transformers, which have excelled in language modeling~\cite{radford2019language,brown2020language} and 3D mesh generation~\cite{siddiqui2024meshgpt,chen2024meshanything}, offer promising solutions for generating variable-length sequences. However, efficiently modeling anime hairstyle within an AR framework remains non-trivial due to the stylistic complexity and geometric variability inherent in such representations. Our anime hairstyle representation is both compact and invertible, enabling efficient learning in auto-regressive generation.

\section{Method}

Our proposed \ourmethod\ is the first comprehensive framework dedicated to 3D anime hairstyle generation. In Sec.~\ref{subsec:data}, we formally define 3D anime hairstyle and describe our dataset construction process. Sec.~\ref{subsec:param} explores our novel hair parameterization approach, while Sec.~\ref{subsec:network} elaborates on the hairstyle transformer architecture and sequence formulation for auto-regressive generation. Finally, Sec.~\ref{subsec:train} outlines the training and inference mechanisms.

\subsection{Definition and Dataset for 3D Anime Hairstyles}
\label{subsec:data}

Anime hairstyles are widely utilized on non-realistic characters in video games, virtual reality, and other digital environments. They differ fundamentally from realistic hair representations. Realistic hair is typically represented as strands, where each individual hair is treated as a strand with positional information but lacking 3D structural details. Additionally, realistic hair models generally maintain a fixed number of strands with predetermined root positions.

\begin{figure}[t]
\centering
\includegraphics[width=0.95\linewidth]{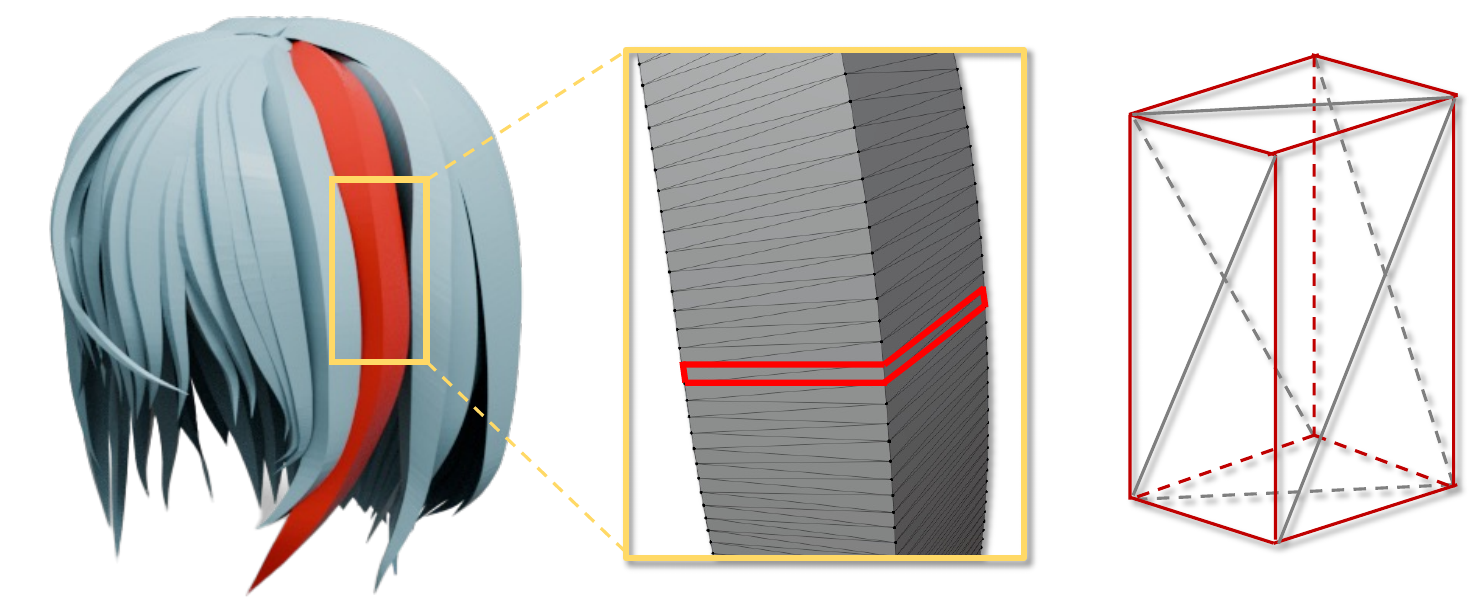}
\caption{Illustration of the Anime hairstyle structure. Left: A complete hairstyle sample. Middle: Repeating units in the highlighted hair card. Right: Mesh connection pattern of a single unit.}
\label{fig:unit}
\end{figure}

In contrast, anime hairstyle consists of multiple independent 3D meshes, each representing a single hair card. These individual hair cards are composed of several small, repeating 3D units connected end-to-end, with variable root positions across the scalp. As illustrated in Fig.~\ref{fig:unit}, the left side displays a complete anime hairstyle, with a highlighted hair card enlarged to reveal its structure. This individual hair consists of multiple quadrilateral pyramids connected sequentially, featuring diamond-shaped top and bottom faces. Within each repeating unit, the mesh edge connections follow consistent and regular patterns, while maintaining connectivity between adjacent units despite variations in size. In the current industry production pipelines and available data, these repeating units are predominantly of two types: diamond-shaped quadrilateral pyramids and triangular pyramids. Each hair card typically comprises 20-60 repeating units, with a complete hairstyle consisting of 25-130 individual hair cards.

Currently there are no existing anime hairstyle datasets available for data-driven learning.
To facilitate the procedural generation and manipulation of anime hairstyles, we have assembled AnimeHair, a large-scale dataset containing 37K distinct anime hairstyles. These samples were derived from publicly available 3D character models on VRoid-Hub. The extraction process involved isolating the hair components and implementing filters to eliminate corner cases or low-quality samples that deviated from standard hair construction principles (such as non-watertight meshes, discontinuous meshes, or arbitrary shapes). Further details of dataset construction and processing are provided in the supplementary material (Sec.~B).

\subsection{Invertible Hair Parameterization}
\label{subsec:param}
To enable autoregressive generation of 3D anime-style hair meshes, it is necessary to compress the original geometry before generation. Directly generating high-resolution meshes is infeasible due to computational constraints: typical hairstyles contain 7K-50K faces, far exceeding the capacity of current autoregressive 3D mesh generation methods, making a compact representation essential.

Given the structured nature of anime hairstyles, a natural approach is to parameterize the repeating units. Upon inspecting template data from existing game assets, we observed that most of the hair cards are constructed using a Bézier surface and parameterized through two attributes: width and thickness. Each hair card is constrained to lie on a Bézier surface defined by a set of control points. The segment between any two consecutive control points can be treated as a repeating unit mentioned above. At each control point, the width and thickness value correspond to either the diagonals of a diamond or the base and height of an isosceles triangle, representing the hair’s cross-sectional dimensions. The thickness direction aligns with the normal of the Bézier surface at that point, while the width direction is computed as the cross product of the surface normal and the tangent vector along the control curve.

A key challenge is determining a minimal yet expressive parameterization of each unit. Directly storing width and thickness directions per cross-section is intuitive but redundant and less learnable. Reconstructing the original Bézier surface from each hair card mesh is computationally expensive, unstable, and ill-posed due to the unknown and variable number of repeating units. This motivates the need for a compact and invertible parameterization scheme.

\begin{figure}[t]
\centering
\includegraphics[width=0.9\linewidth]{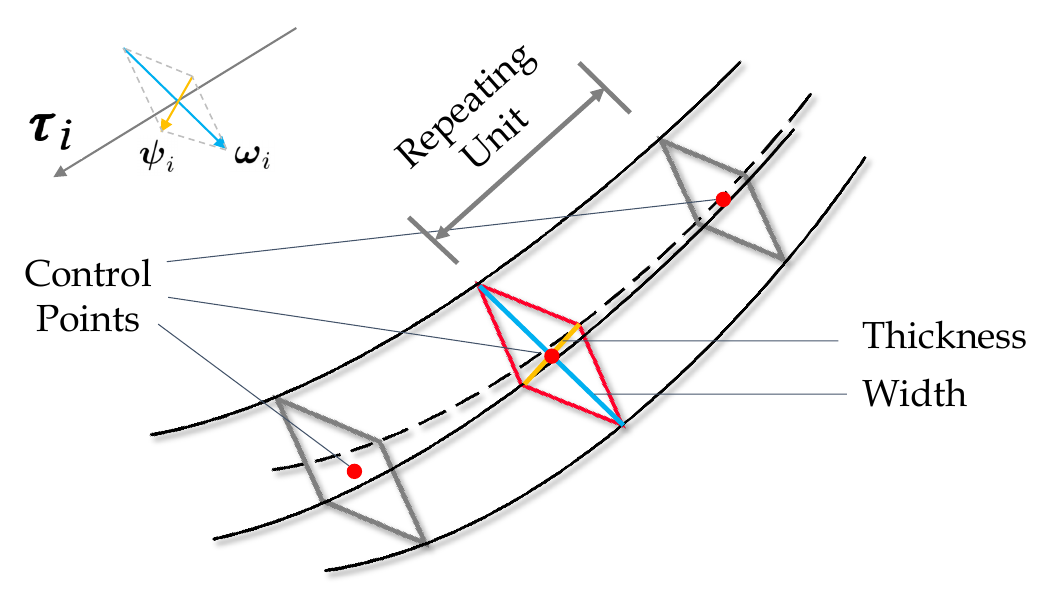}
\caption{Illustration of our hair parameterization. Within a single hair card, each bottom face of a repeating unit is defined by the 3D position of the centered control point, width, and thickness.}
\label{fig:param}

\end{figure}

To address these issues, we design a fully invertible control-point-based parameterization tailored for anime hairstyles. Each hair card is represented by $N$ control points, and each control point is characterized by only 5 float values: 3D position $(x, y, z)$, width $w$, and thickness $t$ (shown in Fig.~\ref{fig:param}). The tangent vector $\boldsymbol{\tau}_i$ at each control point $i$ is estimated using a fourth-order accurate five-point central difference method. Given a control point $i$, its tangent is approximated using the coefficient vector $[-1/12, 2/3, 0, -2/3, 1/12]$ applied to its five nearest neighbors.

Estimating reliable width vectors $\boldsymbol{\omega}_i$ for each control point $i$ by only the control point sequence poses an additional challenge. A common approach is to apply Principal Component Analysis~(PCA) to the neighboring points and use the resulting normal vector as the width direction. However, this method performs poorly in regions where the hair are nearly straight, often producing unstable and noisy results. To address this, we reformulate the problem as an optimization task with two key objectives:

\begin{enumerate}
    \item The projection distances of neighboring points onto the normal plane should be minimized.
    \item Normals at adjacent points should vary smoothly.
\end{enumerate}

Simply calculating initial normal vectors and propagating with Bishop frames~\cite{bergou2008discrete} suffers from precision issues due to the fewer and varied-sized repeating units, as well as the higher discretization characteristics of anime hairstyles. While gradient descent can be used to solve this problem and yields reasonable results, it is computationally expensive and lacks determinism, making it unsuitable for a template-driven generation framework. To improve efficiency and robustness, we reformulate the problem as a least-squares optimization, removing the unit-length constraint on normals and initializing the first normal (e.g., via PCA) to prevent degenerate solutions. This formulation allows us to solve for all normals in a single linear system. For the coordinates $p_i$ of each control point $i$, we compute the difference vectors with its neighbors:
\begin{align}
    d_{i,\text{prev}} = p_i - p_{i-1}, \quad d_{i,\text{next}} = p_i - p_{i+1}
\end{align}
We then form the data matrix:
\begin{align}
    D_i = d_{i,\text{prev}}d_{i,\text{prev}}^T + d_{i,\text{next}}d_{i,\text{next}}^T
\end{align}
The optimization objective consists of two terms. The first encourages the estimated normal $\mathbf{n}_i$ to be orthogonal to the local geometry encoded by $D_i$, minimizing the projection of the surrounding structure onto the normal direction. The second term imposes smoothness by penalizing differences between adjacent normals. The complete objective is given by:
\begin{align}
    \sum_i \left( \mathbf{n}_i^T D_i \mathbf{n}_i \right) + \lambda \sum_i \|\mathbf{n}_i - \mathbf{n}_{i+1}\|^2
    \label{eq:obj}
\end{align}
This formulation leads to a sparse linear system of the form $A \mathbf{n} = 0$, where the matrix $A$ is block tridiagonal. Where each diagonal block is defined as $A[i][i] = D_i + 2\lambda I $, and off-diagonal blocks that connect neighboring normals are $A[i][i+1] = A[i+1][i] = -\lambda I$. To avoid the trivial zero solution, we fix the normal $\mathbf{n}_1$ at the first control point (e.g., via PCA initialization), effectively anchoring the solution. The system is then solved using standard least-squares methods to recover the normal vector $\mathbf{n}_i$ at each control point.

With the system fully defined and the constraint applied, we solve for the entire set of normal vectors using standard least-squares solvers, obtaining a globally consistent and smooth field of normals along the single hair card. The direction vectors of width $\boldsymbol{\omega}_i$ and thickness $\boldsymbol{\psi}_i$ at each control point $i$ are then defined as:
\begin{align}
    \boldsymbol{\omega}_i = \mathbf{n}_i, \quad \boldsymbol{\psi}_i = \mathbf{n}_i \times \boldsymbol{\tau}_i
\end{align}
With the width and thickness direction vectors defined, we can now reconstruct the vertex positions of the bottom face of each repeated hair unit. Taking the diamond-shaped unit as an example, the four corner vertices corresponding to control point $p_i$ are given by $[p_i + \frac{1}{2} \boldsymbol{\omega}_i, p_i - \frac{1}{2} \boldsymbol{\omega}_i, p_i + \frac{1}{2} \boldsymbol{\psi}_i, p_i - \frac{1}{2} \boldsymbol{\psi}_i]$. Using these vertices, we can assemble the geometry of each repeating unit and sequentially reconstruct the complete 3D mesh of a single hair card.

\begin{figure*}[htb]
\centering
\includegraphics[width=0.93\linewidth]{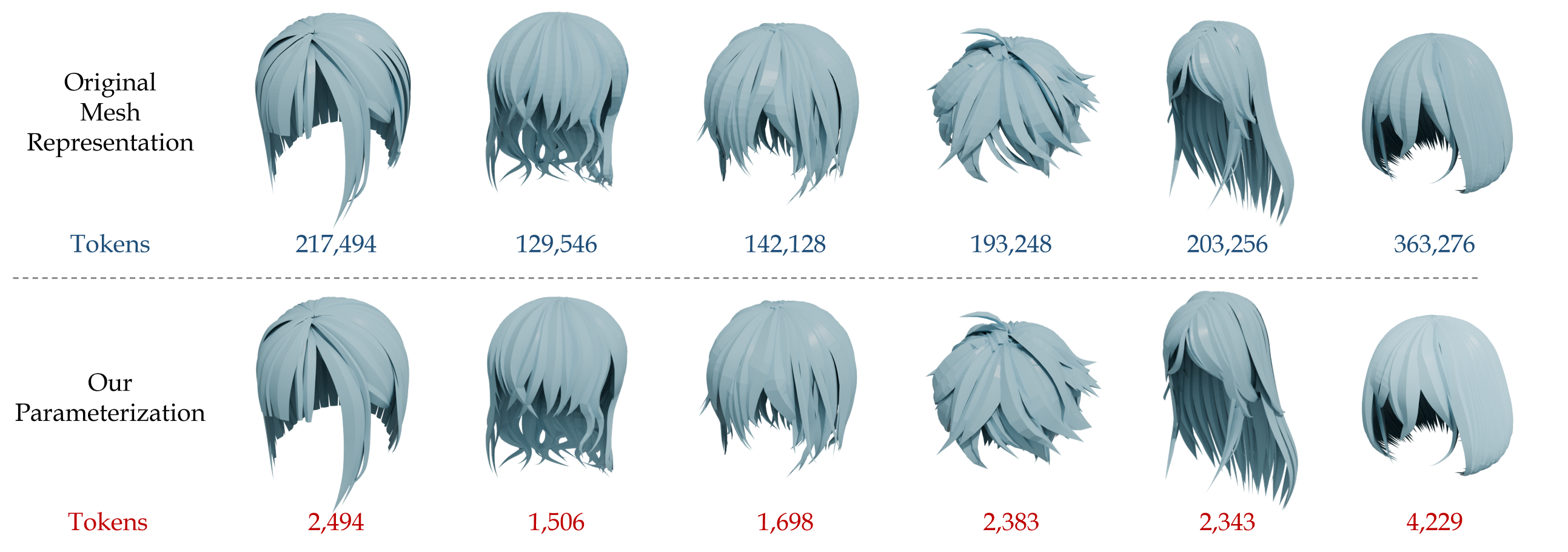}
\caption{Our compact and invertible parameterization achieves over 98\% token compression compared to the original mesh.}
\label{fig:token}
\end{figure*}

At this point, we have established a forward pipeline that reconstructs the full 3D mesh of a single hair card given only $N$ control points, where each point stores five parameters: its 3D position $p = (x, y, z)$, width $w$, and thickness $t$. The inverse process is comparatively straightforward. For each repeating unit, we extract the centroid of its bottom face as the corresponding control point position. The diagonals (in the case of a diamond) or the base and height (in the case of an isosceles triangle) are directly mapped to the width and thickness values. Through this design, we achieve a fully invertible and compact parameterization of anime-style hair geometry, preserving anime styling at <2\% of the original 3D mesh token cost (Fig.~\ref{fig:token}).

\subsection{Auto-Regressive Hairstyle Transformer}
\label{subsec:network}
We now design an efficient and effective network structure with a flexible sequence formulation to implement auto-regressive anime hairstyle generation.

\subsubsection{Network Structure}
Our hairstyle transformer 
takes a condition $\mathcal{C}$ (in our case, a point cloud) as input and outputs a sequence of control points that constitute a hairstyle. The
architecture comprises three primary components: a control point encoder $\mathcal{E}$, a decoder-only transformer $\mathcal{T}$, and a cascaded decoder $\mathcal{D}$. We adopt a piecewise discretization scheme to quantize each control point $i$'s attributes, including position $p_i=(x_i, y_i, z_i)$, width $w_i$, and thickness $t_i$ (detail in supplementary Sec.~C), and treat them as discrete input tokens. These tokens are processed through learnable embeddings $\mathbf{e}_p$, $\mathbf{e}_w$, and $\mathbf{e}_t$, then fed into the linear control point encoder $\mathcal{E}$ to generate control point tokens $h_i$:
\begin{align}
    h_i = \mathcal{E}([\mathbf{e}_p(p_i); \mathbf{e}_w(w_i); \mathbf{e}_t(t_i)])
\end{align}

The decoder-only transformer $\mathcal{T}$ performs next-token prediction at the control point level. Given input condition $\mathcal{C}$ and all previously predicted tokens ${h_1, \dots, h_{i-1}}$, it generates the feature representation $\mathbf{f}_i$ for the next control point:
\begin{align}
    \mathbf{f}_i = \mathcal{T}(\mathcal{C}; h_1, \dots, h_{i-1})
\end{align}
Subsequently, we decode specific attributes from this feature representation. Following previous generation works~\cite{ritchie2019fast,wang2021sceneformer,paschalidou2021atiss, ye2025primitiveanything}, we employ a cascaded decoding structure that explicitly models the dependencies among control point attributes:
\begin{align}
    \hat{p}_i & = \mathcal{D}_p(\mathbf{f}_i), \\
    \hat{w}_i & = \mathcal{D}_w(\mathbf{f}_i, \mathbf{e}_p(\hat{p}_i)), \\
    \hat{t}_i & = \mathcal{D}_t(\mathbf{f}_i, \mathbf{e}_p(\hat{p}_i), \mathbf{e}_w(\hat{w}_i))
\end{align}
Where $\mathcal{D}_p$, $\mathcal{D}_w$, and $\mathcal{D}_t$ represent the position decoder, width decoder, and thickness decoder, respectively. Each decoder operates on the concatenated input of $\mathbf{f}_i$ and the embedded representations of past decoded attributes, outputting probability logits.

\begin{figure}[t]
\centering
\includegraphics[width=1.0\linewidth]{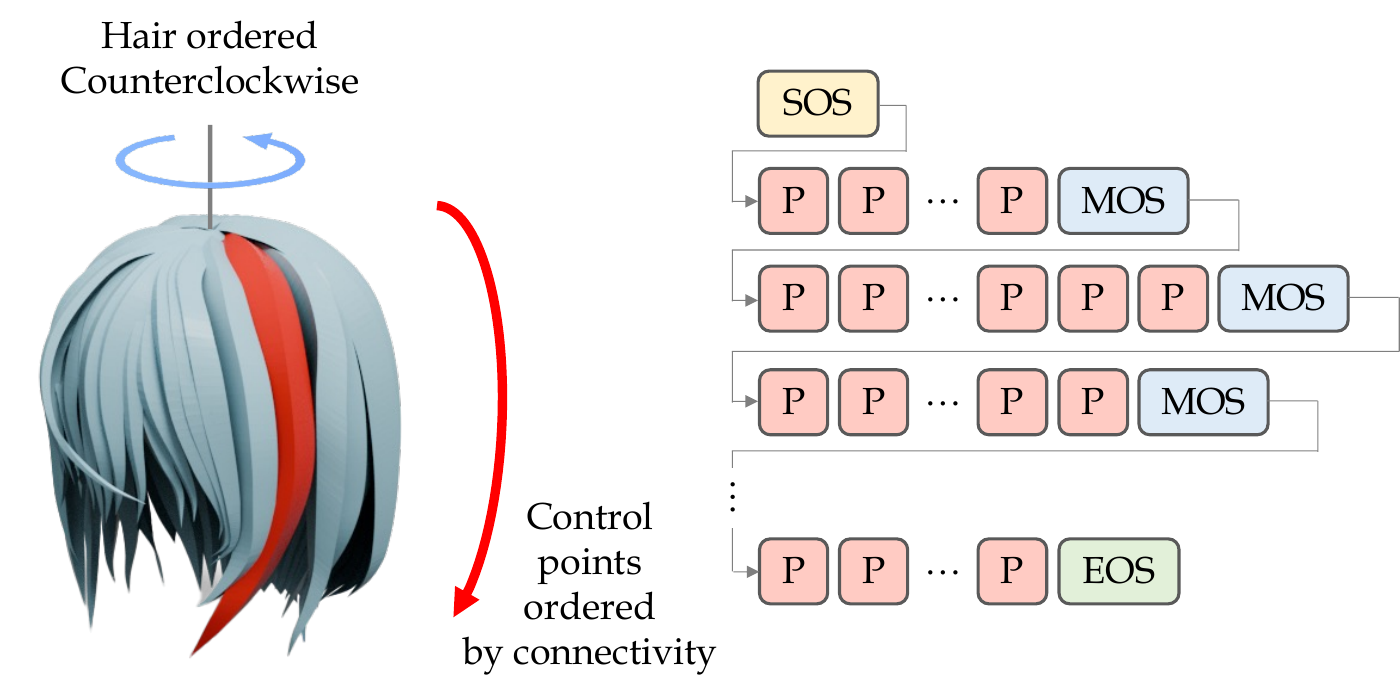}
\caption{Hair sequence construction. Left: Ordering strategy showing counterclockwise arrangement of hair cards around the head and root-to-tip connectivity within each hair. Right: Sequence formulation where each row corresponds to a single hair card terminated by a MOS token.}
\label{fig:seq}
\end{figure}

\subsubsection{Sequence Formulation}
Given that anime hairstyles comprise a variable number of individual hair cards, each with different control point counts and root positions, we have designed a flexible sequence formulation to interpret anime hairstyles as a sequential ``hair language'', enabling autoregressive generation of arbitrary hairstyles (shown in Fig.~\ref{fig:seq}).

At the hairstyle level (the ``sentence'' level of our ``hair language''), we first define a generation order starting from the back of the head. Each individual hair card is ordered based on its average x,z coordinates (with the y-axis pointing upward), arranged counterclockwise when viewed from above. This ordering approach effectively captures dependencies between adjacent hair cards and facilitates continuous, stable generation, as it ensures spatial continuity between consecutive hair cards in the generation sequence.

For the internal structure of each hair card (the ``word'' level within a ``sentence''), rather than ordering control points strictly by position, we arrange them sequentially from root to tip according to their connectivity relationships. This better preserves continuity and respects the physical characteristics of hair. Considering the natural properties of hair, we define the root of a single hair card as the endpoint that lies closer to the top part of the y-axis (vertical axis) passing through the hairstyle's center.

After defining the ordering system, we concatenate all control points into a final sequence. Similar to other autoregressive generation methods, we prepend input condition tokens followed by a start token \text{<SOS>} at the beginning of the sequence and append an end token \text{<EOS>} at the end. Notably, we introduce a special middle token \text{<MOS>} between individual hair cards, which provides the model with explicit signals to transition between different structural components of the hairstyle. We implement two additional decoders, $\mathcal{D}_{mos}$ and $\mathcal{D}_{eos}$, which operate on the feature representation $\mathbf{f}_i$ to determine when to terminate a hair card and the entire hairstyle, respectively.
This sequence formulation enables our model to generate hairstyles with arbitrary numbers of hair cards, each with variable lengths and control point configurations.

\subsection{Training and Inference}
\label{subsec:train}

\subsubsection{Training Objective}
We use the surface point cloud as the input condition $\mathcal{C}$ and employ Michelangelo~\cite{zhao2023michelangelo} encoder to transform it into a fixed-length token sequence. This sequence is prepended with the \text{<SOS>} token. The hairstyle transformer is trained using a next-step prediction objective with the following loss function: $ \mathcal{L}= \mathcal{L}_{ce} + \mathcal{L}_{eos} + \mathcal{L}_{mos} $,
Where $\mathcal{L}_{ce}$ denotes the cross-entropy loss used to supervise the discrete control point attributes $p_i, w_i, t_i$. The terms $\mathcal{L}_{eos}$, $\mathcal{L}_{mos}$ refer to binary cross-entropy losses applied to $\mathcal{D}_{eos}(\mathbf{f}_i)$ and $\mathcal{D}_{mos}(\mathbf{f}_i)$, which guide the predictions for hairstyle-level and hair-card-level termination, respectively.

\subsubsection{Inference}
Our inference begins with the input condition and <SOS> token, from which we auto-regressively generate control point features $\{\mathbf{f}_i\}_{i=1}^n$ and subsequently decode them into their corresponding attributes. When the MOS decoder determines that the current hair card should terminate, the current token is replaced with an MOS token to signal the end of that hair. Additionally, we implement a structural coherence constraint: if a generated control point is the sixth or subsequent point in a hair card and its position deviates significantly from the extrapolated position predicted by a cubic spline fitted to preceding points (we use 0.03 as the threshold), we replace it with an MOS token. This constraint prevents unrealistic geometric discontinuities within individual hair cards.

The generation process continues iteratively until the EOS judgment criterion is satisfied. Specifically, we allow the entire hairstyle to terminate when both the number of generated individual hair cards exceeds 10 and the feature representation $\mathbf{f}_i$ is classified as an EOS token by the decoder $\mathcal{D}_{eos}$, which prevents the early stop of the hairstyle generation.

\section{Experiments}

\begin{figure*}[t]
\centering
\includegraphics[width=1.0\linewidth]{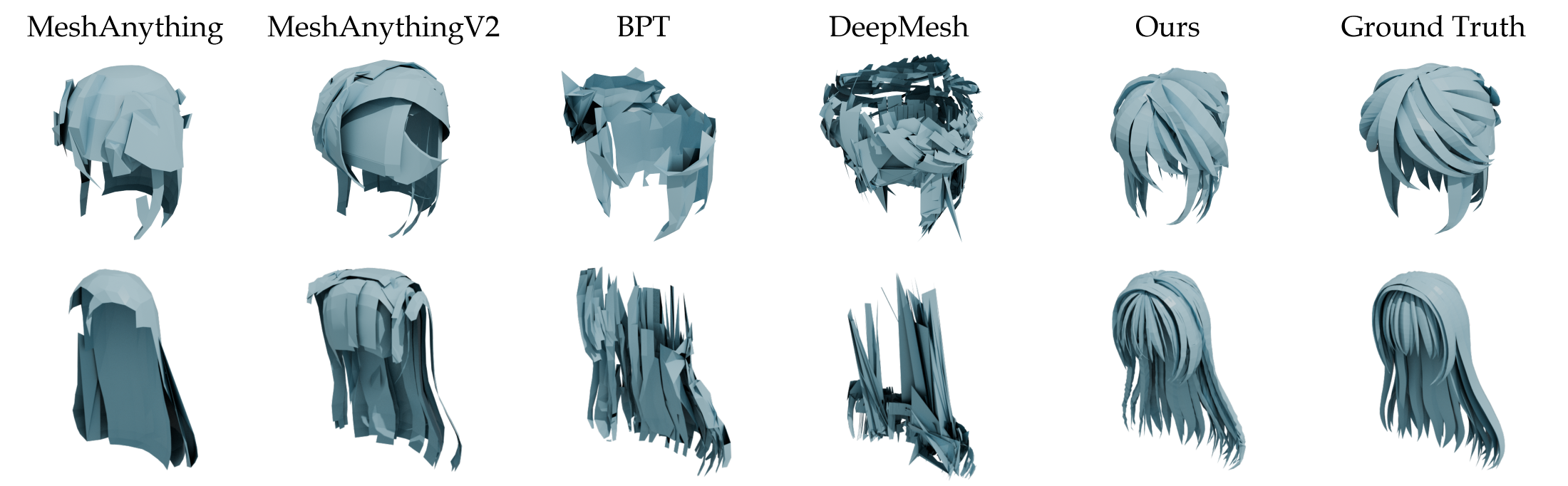}
\caption{Qualitative comparisons with other shape-conditioned 3D mesh generation methods.}
\label{fig:main}
\end{figure*}

\subsection{Experimental Setup}
\subsubsection{Baselines}
We randomly selected 100 hairstyles from the AnimeHair dataset as our test set, ensuring no overlap with the training set.
Given the absence of existing methods specifically designed for the shape-conditioned generation of anime-style hair meshes, we benchmark our approach against state-of-the-art shape-conditioned 3D mesh generation methods, including MeshAnything~\cite{chen2024meshanything}, MeshAnythingV2~\cite{chen2024meshanythingv2}, BPT~\cite{weng2024scaling} and DeepMesh~\cite{zhao2025deepmesh}.

\subsubsection{Metrics}
We uniformly sampled point clouds on the surfaces of predicted anime-style hairstyle mesh predictions and compared them against ground-truth point clouds derived from original reference hairstyle meshes. We employed four distinct quantitative metrics: Chamfer Distance~(CD), Earth Mover's Distance~(EMD), Hausdorff Distance, and Voxel Intersection over Union~(Voxel-IoU).
The Voxel-IoU metric involved a preprocessing step where both predicted and ground-truth point clouds were voxelized at a uniform resolution of $16^3$, and then computing their intersection over union.
To ensure fair comparison given the different normalization approaches used by various methods, we standardize the evaluation process by normalizing the point clouds according to each method's default normalization procedure, and then unnormalizing the generated results using the same method-specific parameters to calculate the metrics.
However, we recognize that geometric metrics alone cannot fully capture cases where a generated hairstyle might cover the surface adequately but exhibit poor morphological characteristics. To further reflect perceptual alignment, we apply fixed lighting conditions to all hairstyles and render each case from eight horizontally equidistant viewpoints. These renderings are then encoded using CLIP~\cite{radford2021learning}, and the average cosine similarity is calculated to quantify perceptual quality.

\subsection{Results and Comparisons}
Our quantitative comparison results are presented in Tab.~\ref{tab:geo}, demonstrating that our method outperforms all other approaches across the evaluated metrics. Most baseline methods struggle to achieve accurate geometric reconstruction of anime-style hairstyles; even MeshAnythingV2, which exhibits more stable performance than other baselines, remains inferior to our approach.

\begin{table}[htbp]
\centering
\caption{Geometric comparison on the AnimeHair test set.}
\resizebox{1.0\linewidth}{!}{
\begin{tabular}{l|c|c|c|c}
\toprule
Method & CD~$\downarrow$ & EMD~$\downarrow$ & Hausdorff~$\downarrow$ & Voxel-IoU~$\uparrow$ \\
\midrule
MeshAnything & 0.0179 & 0.0190 & 0.0584 & 0.7014 \\
MeshAnythingV2 & 0.0118 & 0.0132 & 0.0510 & 0.7503 \\
BPT & 0.0270 & 0.0317 & 0.0767 & 0.6328 \\
DeepMesh & 0.0172 & 0.0198 & 0.0613 & 0.7313 \\
Ours & \textbf{0.0117} & \textbf{0.0128} & \textbf{0.0497} & \textbf{0.7566} \\
\bottomrule
\end{tabular}
}
\label{tab:geo}
\end{table}

The perceptual alignment comparison in Tab.~\ref{tab:clip} reveals a more pronounced advantage for our method, reflecting that other shape-conditioned mesh generation approaches fail to capture the distinctive volumetric structure and stylistic features characteristic of anime hairstyles. These methods typically model hairstyle as a continuous surface rather than as discrete, independently articulated hairs with specific geometric properties.

\begin{table}[htbp]
\centering
\caption{Perceptual alignment comparison on the AnimeHair test set.}
\resizebox{1.0\linewidth}{!}{
\begin{tabular}{l|c|c}
\toprule
Method & CLIP Similarity~$\uparrow$ & Ranking~$\downarrow$ \\
\midrule
MeshAnything~\cite{chen2024meshanything} & 0.8435 & 4  \\
MeshAnythingV2~\cite{chen2024meshanythingv2} & 0.8618 & 2 \\
BPT~\cite{weng2024scaling} & 0.8404 & 5 \\
DeepMesh~\cite{zhao2025deepmesh} & 0.8576 & 3 \\
Ours & \textbf{0.9258} & \textbf{1} \\
\bottomrule
\end{tabular}
}
\label{tab:clip}
\end{table}

This distinction becomes evident in the qualitative comparisons illustrated in Fig.~\ref{fig:main} as well. Other mesh generation methods produce results characterized by incompleteness, lower resolution, and a tendency to clump individual hair cards together. In contrast, our method maintains complete decoupling between different hair cards, resulting in superior human perceptual quality and physical plausibility. Each hair card maintains its distinct shape, flow, and volume while collectively forming a coherent hairstyle that preserves the aesthetic qualities demanded by anime-style representations.

\subsection{Ablation Study}

To validate the effectiveness of our parameterization method and sequence formulation, we conducted a comprehensive ablation study. For parameterization, we examined whether our current parameter set for each control point (xyz coordinates, width, thickness) is optimal.
We introduced an alternative parameterization scheme called {\it Extended Vector Parameterization}, which additionally incorporates directional vectors for width and thickness as prediction parameters, resulting in 11 float values per control point. We also tested a more direct approach called {\it Explicit Vertex Parameterization}, where we generate the hairstyle mesh directly by predicting all the vertices.
Since adopting original mesh tokenization would exceed current token limitations, we simplified the task by pre-specifying vertex ordering and face connectivity patterns, then reconstructing the original mesh and calculating metrics.

\begin{figure*}[htb]
\centering
\includegraphics[width=0.97\linewidth]{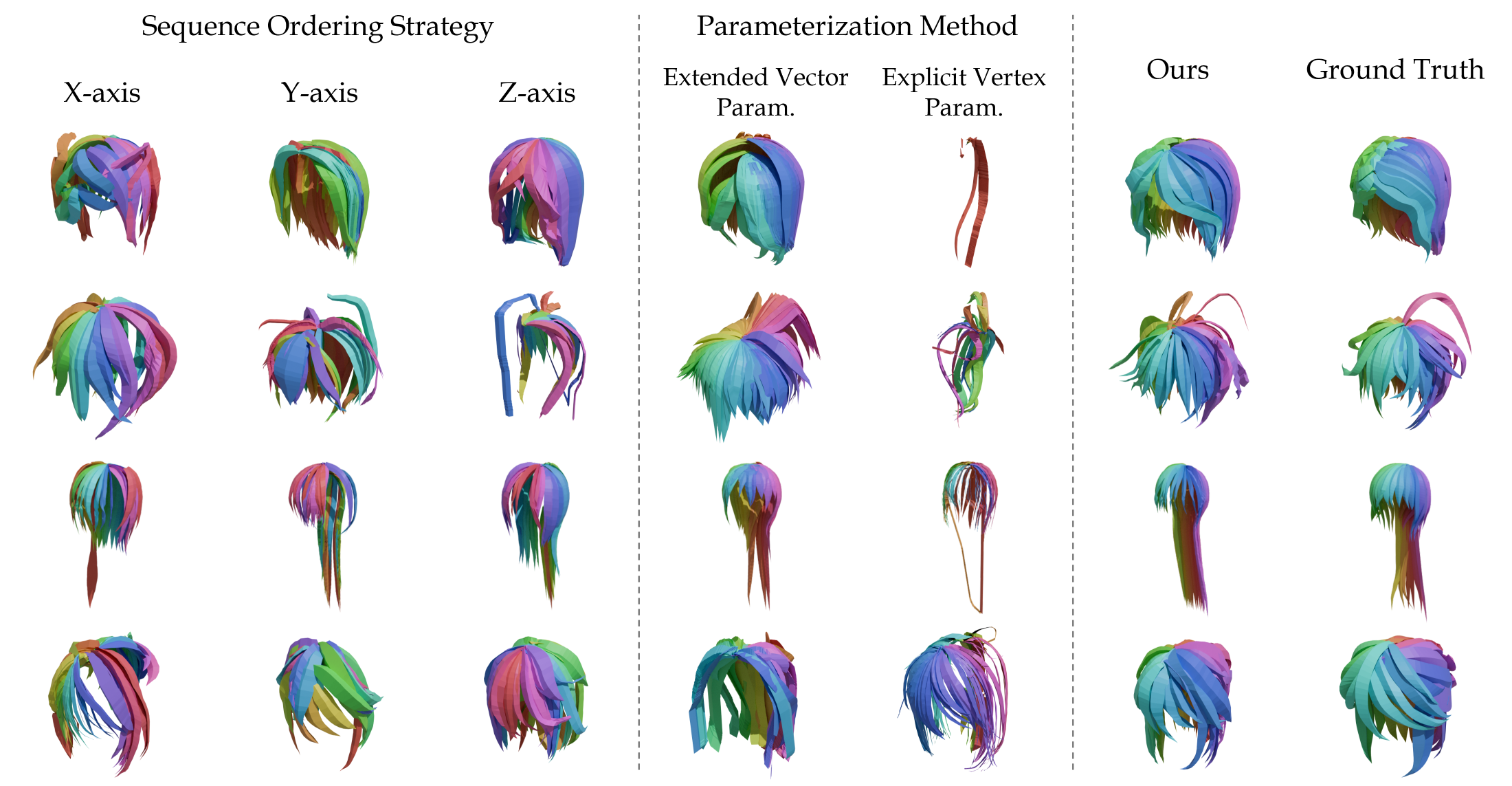}
\caption{Qualitative comparisons of ablation study. Colors indicate different hair strands.}
\label{fig:abl}
\end{figure*}

As shown in Tab.~\ref{tab:abl}, the {\it Extended Vector Parameterization} exhibited some degradation in overall geometric accuracy despite providing more detailed control, highlighting the elegant balance between expressiveness and learnability in our proposed parameterization. The {\it Explicit Vertex Parameterization} demonstrated an even more significant performance decline, underscoring the necessity of transforming mesh prediction into a control-point-based hair language parameterization (as shown in Fig.~\ref{fig:abl}).

For sequence formulation, we compared our counterclockwise hair ordering strategy against common alternatives that sort hair cards based on their average positions along coordinate axes (x-axis, y-axis, z-axis). Tab.~\ref{tab:abl} reveals that y-axis sorting (vertical axis) performed worst, while x-axis and z-axis sorting showed comparable results, though still inferior to our approach. This performance difference arises because counterclockwise hair ordering maximally captures inter-hair patterns and minimizes unpredictable transitions during generation. While x-axis and z-axis sorting maintain some pattern recognition, they frequently introduce spatial discontinuities (alternating between opposite sides of the head), whereas y-axis sorting reveals the least consistent patterns among hair cards.

\begin{figure*}[t]
\centering
\includegraphics[width=0.97\linewidth]{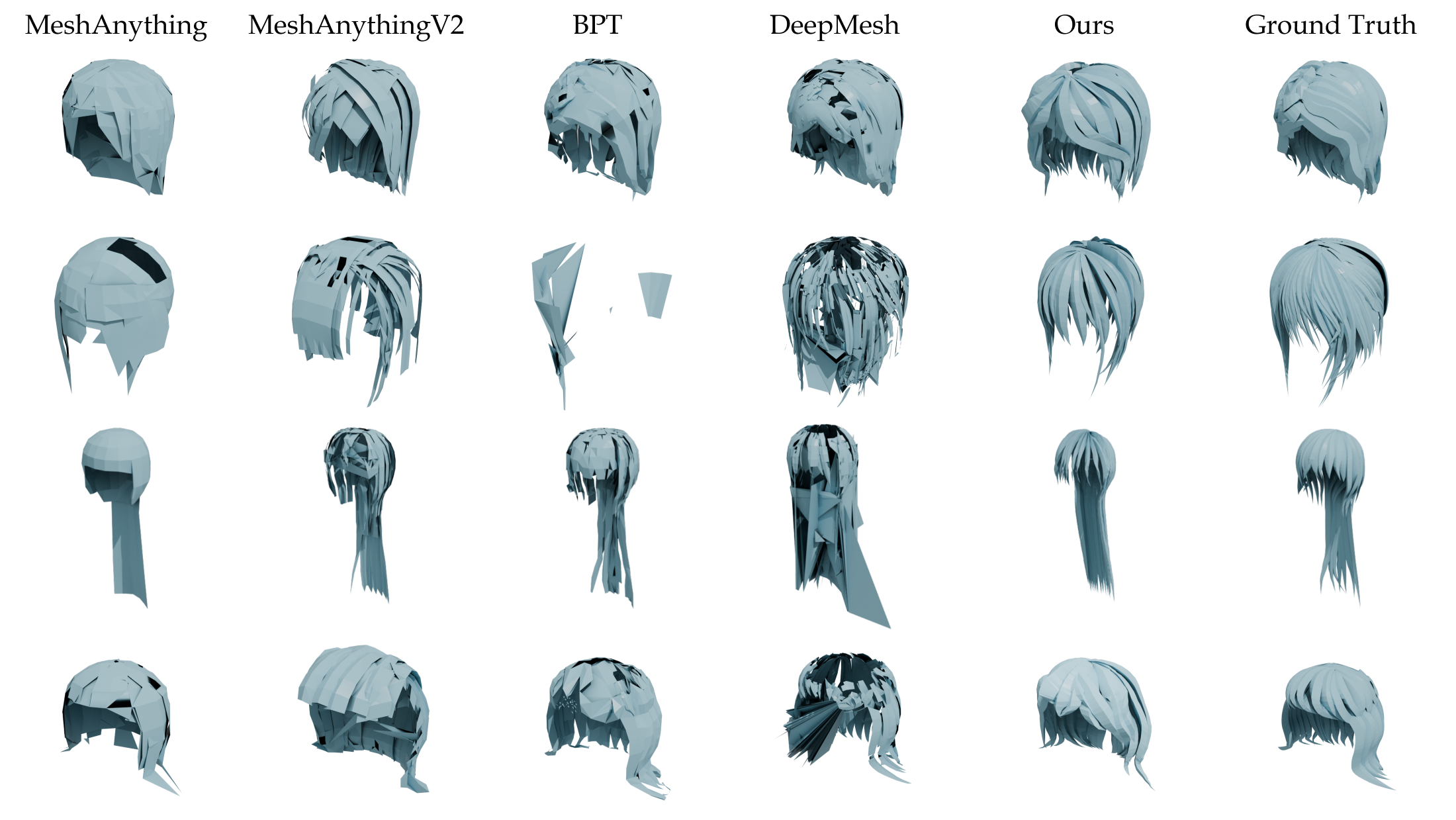}
\caption{More qualitative comparisons with other shape-conditioned 3D mesh generation methods.}
\label{fig:main2}
\end{figure*}

\begin{figure*}[t]
\centering
\vspace{-5pt}
\includegraphics[width=0.88\linewidth]{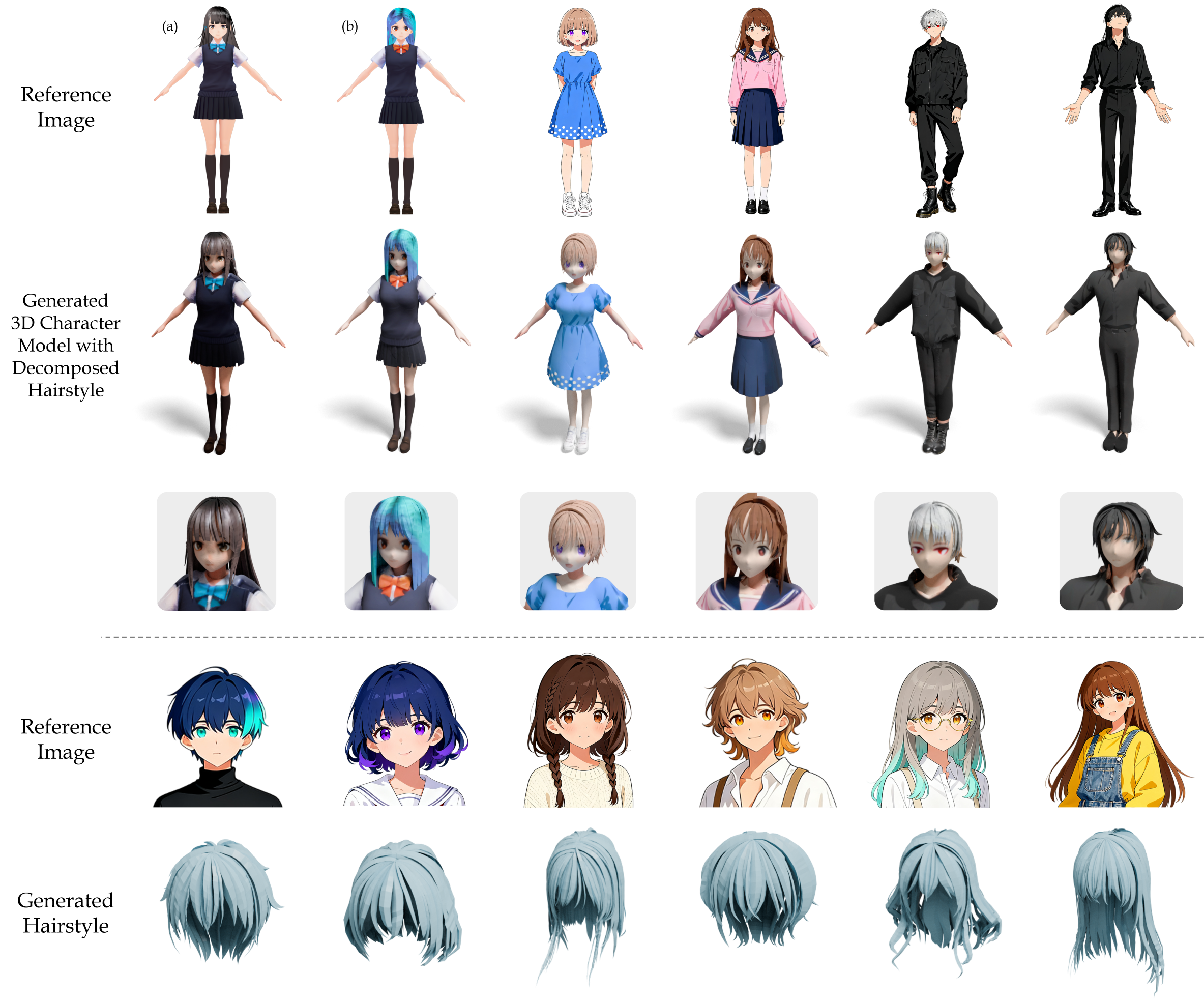}
\vspace{-12pt}
\caption{Our framework supports flexible 3D anime hairstyle generation through various input conditions. The images in (a) and (b) are adapted from publicly available models on VRoid-Hub~\cite{vroidHub}, while the remaining images are AI-generated.}
\label{fig:app}
\end{figure*}

\begin{table}[!t]
\centering
\caption{Ablation study on sequence ordering and parameterization.}
\vspace{-3pt}
\resizebox{1.0\linewidth}{!}{
\begin{tabular}{l|c|c|c|c}
\toprule
Method & CD~$\downarrow$ & EMD~$\downarrow$ & Hausdorff~$\downarrow$ & Voxel-IoU~$\uparrow$ \\
\midrule
\multicolumn{5}{c}{Sequence Ordering Strategy} \\
\midrule
X-Axis Sorting     & 0.0127 & 0.0140 & 0.0531 & 0.7274 \\
Y-Axis Sorting     & 0.0149 & 0.0154 & 0.0521 & 0.6832 \\
Z-Axis Sorting     & 0.0122 & 0.0133 & 0.0527 & 0.7402 \\
Counterclockwise (Ours) & \textbf{0.0117} & \textbf{0.0128} & \textbf{0.0497} & \textbf{0.7566} \\
\midrule
\multicolumn{5}{c}{Parameterization Method} \\
\midrule
Extended Vector Param.   & 0.0129 & 0.0140 & 0.0526 & 0.7411 \\
Explicit Vertex Param.   & 0.0225 & 0.0233 & 0.0857 & 0.6235 \\
Ours & \textbf{0.0117} & \textbf{0.0128} & \textbf{0.0497} & \textbf{0.7566} \\
\bottomrule
\end{tabular}
}
\label{tab:abl}
\end{table}

Given that our method enables decoupled hair representation, we can compute geometric metrics at the hair card level to better reflect the effectiveness of our approach. For each hair card in the prediction, we sample point clouds and identify the hair in the ground truth with the minimum Chamfer Distance; similarly, for each hair card in the ground truth, we find its closest counterpart in the prediction. We then calculate the average of all metrics across these matched pairs and aggregate them over the entire hairstyle, as shown in Tab.~\ref{tab:ablhair}.
The results indicate that our scheme also exhibits relatively strong coherence and accuracy at hair-card level,
validating our approach's ability to generate not only globally plausible hairstyles, but also individually well-formed hairs.

\begin{table}[!t]
\centering
\caption{Ablation study on sequence ordering and parameterization approaches (hair card level). The best metric is highlighted in bold, and the second-best metric is underlined.}
\vspace{-3pt}
\resizebox{1.0\linewidth}{!}{
\begin{tabular}{l|c|c|c|c}
\toprule
Method & CD~$\downarrow$ & EMD~$\downarrow$ & Hausdorff~$\downarrow$ & Voxel-IoU~$\uparrow$ \\
\midrule
\multicolumn{5}{c}{Sequence Ordering Strategy} \\
\midrule
X-Axis Sorting     & \underline{0.0111} & \underline{0.0092} & 0.0289 & 0.6338 \\
Y-Axis Sorting     & 0.0136 & 0.0110 & 0.0320 & 0.6021 \\
Z-Axis Sorting     & 0.0114 & 0.0092 & 0.0306 & 0.6308 \\
Counterclockwise (Ours)    & \textbf{0.0105} & \textbf{0.0085} & \underline{0.0289} & \underline{0.6408} \\
\midrule
\multicolumn{5}{c}{Parameterization Method} \\
\midrule
Extended Vector Param.   & 0.0117 & 0.0094 & \textbf{0.0283} & \textbf{0.6576} \\
Explicit Vertex Param.   & 0.0149 & 0.0115 & 0.0317 & 0.6141 \\
Ours    & \textbf{0.0105} & \textbf{0.0085} & \underline{0.0289} & \underline{0.6408} \\
\bottomrule
\end{tabular}
}
\label{tab:ablhair}
\end{table}

\subsection{Applications}

Our framework supports flexible 3D anime hairstyle generation through various input conditions. As illustrated in Fig.~\ref{fig:app}, our approach can be integrated with StdGEN~\cite{he2024stdgen}, transforming its non-decomposable hair outputs into ready-to-use hairstyles that can be assembled with the original results to create 3D character models more suitable for practical applications. Additionally, our framework supports image-conditioned hairstyle generation by re-training with DINO image features via cross-attention, allowing users to specify desired hairstyle characteristics through reference images. Details are in the supplementary material (Sec.~C).

\subsection{Comparison with Realistic Hair Methods}

\begin{figure}[t]
\centering
\includegraphics[width=1\linewidth]{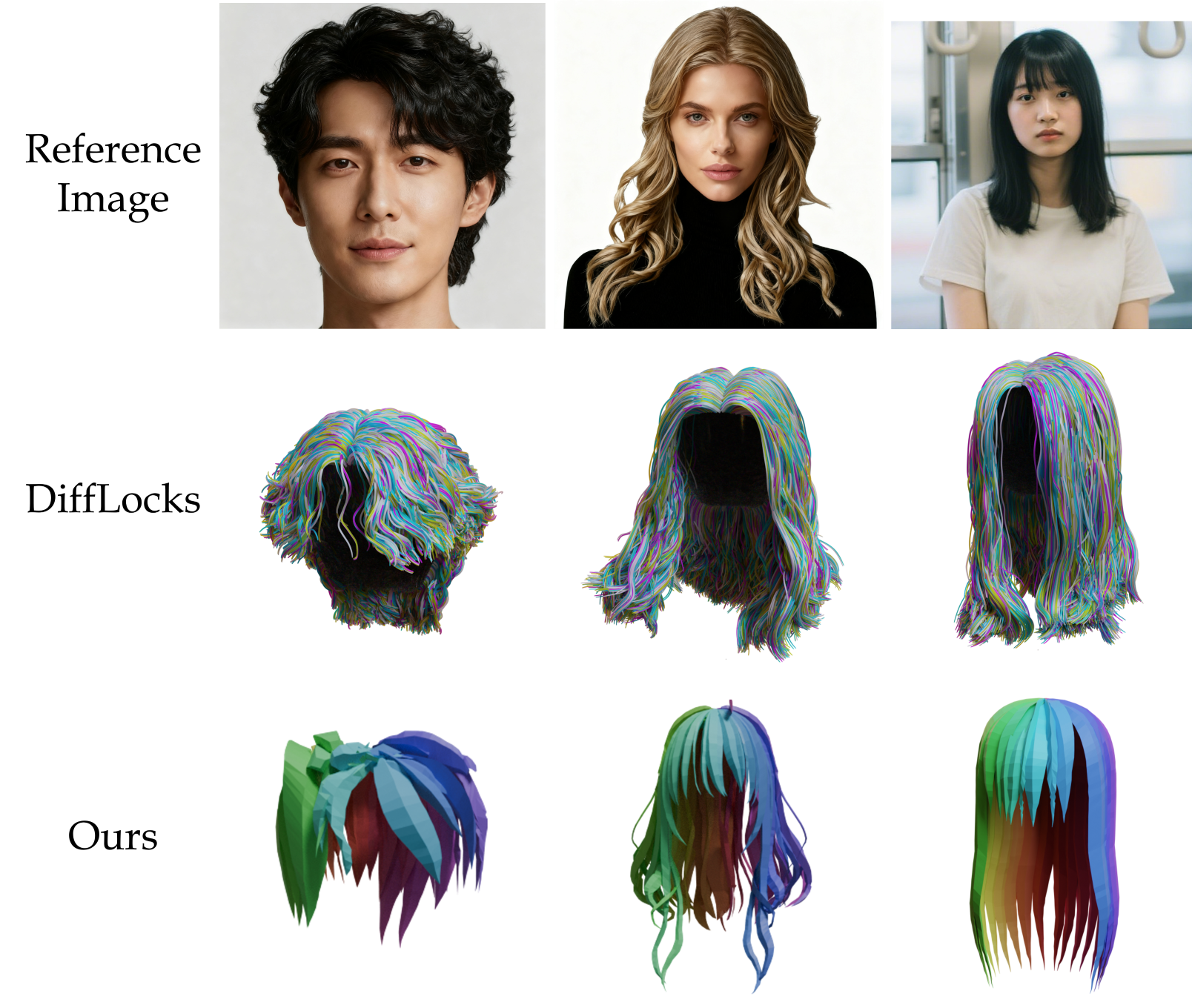}
\caption{Qualitative comparison between DiffLocks (realistic hair method) and our approach on image-conditioned 3D hairstyle generation. The images used for this comparison are AI-generated.}
\label{fig:real}
\end{figure}
To demonstrate the fundamental differences between realistic~\cite{zheng2023hairstep,rosu2025difflocks} and anime hairstyle generation, we conducted a qualitative comparison between image-conditioned methods for realistic hair generation (specifically DiffLocks~\cite{rosu2025difflocks}) and our proposed anime-focused approach. The images used for this comparison are generated by AI and sourced from the AI-generated content website Doubao~\cite{Doubao}, to avoid the copyright and privacy complexities of using real photos. The prompts are available upon request.
As illustrated in Fig.~\ref{fig:real}, the comparison reveals domain-specific limitations when applying methods across different hairstyle domains. Anime hairstyles typically lack realistic elements such as permed hairs, upward-styled curls, and long side-swept hair arrangements. This data distribution gap results in noticeable discrepancies between our predicted 3D hair geometry and the input realistic hair images, as demonstrated in the left and middle cases of Fig.~\ref{fig:real}, while DiffLocks shows better performance.
Conversely, while realistic hair generation methods struggle to accurately predict bangs that fall across the forehead area (even for realistic style input), our anime-specialized approach successfully handles such hairstyles, as shown in the rightmost case. This contrast further underscores the fundamental data distribution differences between realistic and anime hair domains.

\section{Conclusion}
We present \ourmethod, the first framework for 3D anime hairstyle generation, featuring a compact and invertible control point-based parameterization that efficiently represents complex strand geometries. Our specialized hairstyle transformer enables autoregressive generation of detailed hairstyles with individually articulated hair cards. 
Trained on our curated AnimeHair dataset of 37K hairstyles, experiments show significant improvements in geometric accuracy and perceptual quality over existing methods.

\begin{acks}
    This work was partially supported by the Natural Science Foundation of China (62461160309) and NSFC-RGC Joint Research Scheme (N\_HKU705/24).
\end{acks}

\bibliographystyle{ACM-Reference-Format}
\bibliography{bib}


\begin{thebibliography}{63}


\ifx \showCODEN    \undefined \def \showCODEN     #1{\unskip}     \fi
\ifx \showISBNx    \undefined \def \showISBNx     #1{\unskip}     \fi
\ifx \showISBNxiii \undefined \def \showISBNxiii  #1{\unskip}     \fi
\ifx \showISSN     \undefined \def \showISSN      #1{\unskip}     \fi
\ifx \showLCCN     \undefined \def \showLCCN      #1{\unskip}     \fi
\ifx \shownote     \undefined \def \shownote      #1{#1}          \fi
\ifx \showarticletitle \undefined \def \showarticletitle #1{#1}   \fi
\ifx \showURL      \undefined \def \showURL       {\relax}        \fi
\providecommand\bibfield[2]{#2}
\providecommand\bibinfo[2]{#2}
\providecommand\natexlab[1]{#1}
\providecommand\showeprint[2][]{arXiv:#2}

\bibitem[Bergou et~al\mbox{.}(2008)]%
        {bergou2008discrete}
\bibfield{author}{\bibinfo{person}{Mikl{\'o}s Bergou}, \bibinfo{person}{Max Wardetzky}, \bibinfo{person}{Stephen Robinson}, \bibinfo{person}{Basile Audoly}, {and} \bibinfo{person}{Eitan Grinspun}.} \bibinfo{year}{2008}\natexlab{}.
\newblock \showarticletitle{Discrete elastic rods}.
\newblock In \bibinfo{booktitle}{\emph{ACM SIGGRAPH 2008 papers}}. \bibinfo{pages}{1--12}.
\newblock


\bibitem[Brown et~al\mbox{.}(2020)]%
        {brown2020language}
\bibfield{author}{\bibinfo{person}{Tom Brown}, \bibinfo{person}{Benjamin Mann}, \bibinfo{person}{Nick Ryder}, \bibinfo{person}{Melanie Subbiah}, \bibinfo{person}{Jared~D Kaplan}, \bibinfo{person}{Prafulla Dhariwal}, \bibinfo{person}{Arvind Neelakantan}, \bibinfo{person}{Pranav Shyam}, \bibinfo{person}{Girish Sastry}, \bibinfo{person}{Amanda Askell}, {et~al\mbox{.}}} \bibinfo{year}{2020}\natexlab{}.
\newblock \showarticletitle{Language models are few-shot learners}.
\newblock \bibinfo{journal}{\emph{Advances in neural information processing systems}}  \bibinfo{volume}{33} (\bibinfo{year}{2020}), \bibinfo{pages}{1877--1901}.
\newblock


\bibitem[Caron et~al\mbox{.}(2021)]%
        {caron2021emerging}
\bibfield{author}{\bibinfo{person}{Mathilde Caron}, \bibinfo{person}{Hugo Touvron}, \bibinfo{person}{Ishan Misra}, \bibinfo{person}{Herv{\'e} J{\'e}gou}, \bibinfo{person}{Julien Mairal}, \bibinfo{person}{Piotr Bojanowski}, {and} \bibinfo{person}{Armand Joulin}.} \bibinfo{year}{2021}\natexlab{}.
\newblock \showarticletitle{Emerging properties in self-supervised vision transformers}. In \bibinfo{booktitle}{\emph{Proceedings of the IEEE/CVF international conference on computer vision}}. \bibinfo{pages}{9650--9660}.
\newblock


\bibitem[Chen et~al\mbox{.}(1999)]%
        {chen1999asystem}
\bibfield{author}{\bibinfo{person}{Lieu-Hen Chen}, \bibinfo{person}{Santi Saeyor}, \bibinfo{person}{Hiroshi Dohi}, {and} \bibinfo{person}{Mitsuru Ishizuka}.} \bibinfo{year}{1999}\natexlab{}.
\newblock \showarticletitle{A system of 3D hair style synthesis based on the wisp model}.
\newblock \bibinfo{journal}{\emph{The Visual Computer}} \bibinfo{volume}{15}, \bibinfo{number}{4} (\bibinfo{date}{Jul} \bibinfo{year}{1999}), \bibinfo{pages}{159–170}.
\newblock
\href{https://doi.org/10.1007/s003710050169}{doi:\nolinkurl{10.1007/s003710050169}}


\bibitem[Chen et~al\mbox{.}(2025)]%
        {chen2025dora}
\bibfield{author}{\bibinfo{person}{Rui Chen}, \bibinfo{person}{Jianfeng Zhang}, \bibinfo{person}{Yixun Liang}, \bibinfo{person}{Guan Luo}, \bibinfo{person}{Weiyu Li}, \bibinfo{person}{Jiarui Liu}, \bibinfo{person}{Xiu Li}, \bibinfo{person}{Xiaoxiao Long}, \bibinfo{person}{Jiashi Feng}, {and} \bibinfo{person}{Ping Tan}.} \bibinfo{year}{2025}\natexlab{}.
\newblock \showarticletitle{Dora: Sampling and benchmarking for 3d shape variational auto-encoders}. In \bibinfo{booktitle}{\emph{Proceedings of the Computer Vision and Pattern Recognition Conference}}. \bibinfo{pages}{16251--16261}.
\newblock


\bibitem[Chen et~al\mbox{.}(2023)]%
        {chen2023panic}
\bibfield{author}{\bibinfo{person}{Shuhong Chen}, \bibinfo{person}{Kevin Zhang}, \bibinfo{person}{Yichun Shi}, \bibinfo{person}{Heng Wang}, \bibinfo{person}{Yiheng Zhu}, \bibinfo{person}{Guoxian Song}, \bibinfo{person}{Sizhe An}, \bibinfo{person}{Janus Kristjansson}, \bibinfo{person}{Xiao Yang}, {and} \bibinfo{person}{Matthias Zwicker}.} \bibinfo{year}{2023}\natexlab{}.
\newblock \showarticletitle{Panic-3d: Stylized single-view 3d reconstruction from portraits of anime characters}. In \bibinfo{booktitle}{\emph{Proceedings of the IEEE/CVF Conference on Computer Vision and Pattern Recognition}}. \bibinfo{pages}{21068--21077}.
\newblock


\bibitem[Chen et~al\mbox{.}(2024a)]%
        {chen2024meshanything}
\bibfield{author}{\bibinfo{person}{Yiwen Chen}, \bibinfo{person}{Tong He}, \bibinfo{person}{Di Huang}, \bibinfo{person}{Weicai Ye}, \bibinfo{person}{Sijin Chen}, \bibinfo{person}{Jiaxiang Tang}, \bibinfo{person}{Xin Chen}, \bibinfo{person}{Zhongang Cai}, \bibinfo{person}{Lei Yang}, \bibinfo{person}{Gang Yu}, {et~al\mbox{.}}} \bibinfo{year}{2024}\natexlab{a}.
\newblock \showarticletitle{MeshAnything: Artist-Created Mesh Generation with Autoregressive Transformers}.
\newblock \bibinfo{journal}{\emph{arXiv preprint arXiv:2406.10163}} (\bibinfo{year}{2024}).
\newblock


\bibitem[Chen et~al\mbox{.}(2024b)]%
        {chen2024meshanythingv2}
\bibfield{author}{\bibinfo{person}{Yiwen Chen}, \bibinfo{person}{Yikai Wang}, \bibinfo{person}{Yihao Luo}, \bibinfo{person}{Zhengyi Wang}, \bibinfo{person}{Zilong Chen}, \bibinfo{person}{Jun Zhu}, \bibinfo{person}{Chi Zhang}, {and} \bibinfo{person}{Guosheng Lin}.} \bibinfo{year}{2024}\natexlab{b}.
\newblock \showarticletitle{Meshanything v2: Artist-created mesh generation with adjacent mesh tokenization}.
\newblock \bibinfo{journal}{\emph{arXiv preprint arXiv:2408.02555}} (\bibinfo{year}{2024}).
\newblock


\bibitem[Doubao(2025)]%
        {Doubao}
\bibfield{author}{\bibinfo{person}{Bytedance Doubao}.} \bibinfo{year}{2025}\natexlab{}.
\newblock \bibinfo{title}{Doubao}.
\newblock
\urldef\tempurl%
\url{https://doubao.com/}
\showURL{%
\tempurl}


\bibitem[Gafni et~al\mbox{.}(2021)]%
        {gafni2021dynamic}
\bibfield{author}{\bibinfo{person}{Guy Gafni}, \bibinfo{person}{Justus Thies}, \bibinfo{person}{Michael Zollhofer}, {and} \bibinfo{person}{Matthias Nie{\ss}ner}.} \bibinfo{year}{2021}\natexlab{}.
\newblock \showarticletitle{Dynamic neural radiance fields for monocular 4d facial avatar reconstruction}. In \bibinfo{booktitle}{\emph{Proceedings of the IEEE/CVF Conference on Computer Vision and Pattern Recognition}}. \bibinfo{pages}{8649--8658}.
\newblock


\bibitem[Giebenhain et~al\mbox{.}(2023)]%
        {giebenhain2023learning}
\bibfield{author}{\bibinfo{person}{Simon Giebenhain}, \bibinfo{person}{Tobias Kirschstein}, \bibinfo{person}{Markos Georgopoulos}, \bibinfo{person}{Martin R{\"u}nz}, \bibinfo{person}{Lourdes Agapito}, {and} \bibinfo{person}{Matthias Nie{\ss}ner}.} \bibinfo{year}{2023}\natexlab{}.
\newblock \showarticletitle{Learning neural parametric head models}. In \bibinfo{booktitle}{\emph{Proceedings of the IEEE/CVF Conference on Computer Vision and Pattern Recognition}}. \bibinfo{pages}{21003--21012}.
\newblock


\bibitem[He et~al\mbox{.}(2024a)]%
        {he2024perm}
\bibfield{author}{\bibinfo{person}{Chengan He}, \bibinfo{person}{Xin Sun}, \bibinfo{person}{Zhixin Shu}, \bibinfo{person}{Fujun Luan}, \bibinfo{person}{S{\"o}ren Pirk}, \bibinfo{person}{Jorge Alejandro~Amador Herrera}, \bibinfo{person}{Dominik~L Michels}, \bibinfo{person}{Tuanfeng~Y Wang}, \bibinfo{person}{Meng Zhang}, \bibinfo{person}{Holly Rushmeier}, {et~al\mbox{.}}} \bibinfo{year}{2024}\natexlab{a}.
\newblock \showarticletitle{Perm: A parametric representation for multi-style 3d hair modeling}.
\newblock \bibinfo{journal}{\emph{arXiv preprint arXiv:2407.19451}} (\bibinfo{year}{2024}).
\newblock


\bibitem[He et~al\mbox{.}(2024b)]%
        {he2024stdgen}
\bibfield{author}{\bibinfo{person}{Yuze He}, \bibinfo{person}{Yanning Zhou}, \bibinfo{person}{Wang Zhao}, \bibinfo{person}{Zhongkai Wu}, \bibinfo{person}{Kaiwen Xiao}, \bibinfo{person}{Wei Yang}, \bibinfo{person}{Yong-Jin Liu}, {and} \bibinfo{person}{Xiao Han}.} \bibinfo{year}{2024}\natexlab{b}.
\newblock \showarticletitle{StdGEN: Semantic-Decomposed 3D Character Generation from Single Images}.
\newblock \bibinfo{journal}{\emph{arXiv preprint arXiv:2411.05738}} (\bibinfo{year}{2024}).
\newblock


\bibitem[Hong et~al\mbox{.}(2022)]%
        {hong2022headnerf}
\bibfield{author}{\bibinfo{person}{Yang Hong}, \bibinfo{person}{Bo Peng}, \bibinfo{person}{Haiyao Xiao}, \bibinfo{person}{Ligang Liu}, {and} \bibinfo{person}{Juyong Zhang}.} \bibinfo{year}{2022}\natexlab{}.
\newblock \showarticletitle{Headnerf: A real-time nerf-based parametric head model}. In \bibinfo{booktitle}{\emph{Proceedings of the IEEE/CVF Conference on Computer Vision and Pattern Recognition}}. \bibinfo{pages}{20374--20384}.
\newblock


\bibitem[Kim and Neumann(2002)]%
        {kim2002iteractive}
\bibfield{author}{\bibinfo{person}{Tae-Yong Kim} {and} \bibinfo{person}{Ulrich Neumann}.} \bibinfo{year}{2002}\natexlab{}.
\newblock \showarticletitle{Interactive multiresolution hair modeling and editing}.
\newblock \bibinfo{journal}{\emph{ACM Transactions on Graphics}} (\bibinfo{date}{Jul} \bibinfo{year}{2002}), \bibinfo{pages}{620–629}.
\newblock
\href{https://doi.org/10.1145/566654.566627}{doi:\nolinkurl{10.1145/566654.566627}}


\bibitem[Kirillov et~al\mbox{.}(2023)]%
        {kirillov2023segany}
\bibfield{author}{\bibinfo{person}{Alexander Kirillov}, \bibinfo{person}{Eric Mintun}, \bibinfo{person}{Nikhila Ravi}, \bibinfo{person}{Hanzi Mao}, \bibinfo{person}{Chloe Rolland}, \bibinfo{person}{Laura Gustafson}, \bibinfo{person}{Tete Xiao}, \bibinfo{person}{Spencer Whitehead}, \bibinfo{person}{Alexander~C. Berg}, \bibinfo{person}{Wan-Yen Lo}, \bibinfo{person}{Piotr Doll{\'a}r}, {and} \bibinfo{person}{Ross Girshick}.} \bibinfo{year}{2023}\natexlab{}.
\newblock \showarticletitle{Segment Anything}.
\newblock \bibinfo{journal}{\emph{arXiv:2304.02643}} (\bibinfo{year}{2023}).
\newblock


\bibitem[Koh and Huang(2001)]%
        {koh2001simple}
\bibfield{author}{\bibinfo{person}{Chuan~Koon Koh} {and} \bibinfo{person}{Zhiyong Huang}.} \bibinfo{year}{2001}\natexlab{}.
\newblock \showarticletitle{A simple physics model to animate human hair modeled in 2D strips in real time}. In \bibinfo{booktitle}{\emph{Computer Animation and Simulation 2001: Proceedings of the Eurographics Workshop in Manchester, UK, September 2--3, 2001}}. Springer, \bibinfo{pages}{127--138}.
\newblock


\bibitem[Liang and Huang(2003)]%
        {LiangH03}
\bibfield{author}{\bibinfo{person}{Wenqi Liang} {and} \bibinfo{person}{Zhiyong Huang}.} \bibinfo{year}{2003}\natexlab{}.
\newblock \showarticletitle{An Enhanced Framework for Real-Time Hair Animation}. In \bibinfo{booktitle}{\emph{PG}}. \bibinfo{pages}{467--471}.
\newblock
\urldef\tempurl%
\url{https://doi.org/10.1109/PCCGA.2003.1238296}
\showURL{%
\tempurl}


\bibitem[Liu et~al\mbox{.}(2013)]%
        {liu2013stereo}
\bibfield{author}{\bibinfo{person}{Xueting Liu}, \bibinfo{person}{Xiangyu Mao}, \bibinfo{person}{Xuan Yang}, \bibinfo{person}{Linling Zhang}, {and} \bibinfo{person}{Tien-Tsin Wong}.} \bibinfo{year}{2013}\natexlab{}.
\newblock \showarticletitle{Stereoscopizing cel animations}.
\newblock  (\bibinfo{date}{Nov} \bibinfo{year}{2013}).
\newblock


\bibitem[Long et~al\mbox{.}(2025)]%
        {long2025tangled}
\bibfield{author}{\bibinfo{person}{Pengyu Long}, \bibinfo{person}{Zijun Zhao}, \bibinfo{person}{Min Ouyang}, \bibinfo{person}{Qingcheng Zhao}, \bibinfo{person}{Qixuan Zhang}, \bibinfo{person}{Wei Yang}, \bibinfo{person}{Lan Xu}, {and} \bibinfo{person}{Jingyi Yu}.} \bibinfo{year}{2025}\natexlab{}.
\newblock \showarticletitle{TANGLED: Generating 3D Hair Strands from Images with Arbitrary Styles and Viewpoints}.
\newblock \bibinfo{journal}{\emph{arXiv preprint arXiv:2502.06392}} (\bibinfo{year}{2025}).
\newblock


\bibitem[Luo et~al\mbox{.}(2024)]%
        {luo2024gaussianhair}
\bibfield{author}{\bibinfo{person}{Haimin Luo}, \bibinfo{person}{Min Ouyang}, \bibinfo{person}{Zijun Zhao}, \bibinfo{person}{Suyi Jiang}, \bibinfo{person}{Longwen Zhang}, \bibinfo{person}{Qixuan Zhang}, \bibinfo{person}{Wei Yang}, \bibinfo{person}{Lan Xu}, {and} \bibinfo{person}{Jingyi Yu}.} \bibinfo{year}{2024}\natexlab{}.
\newblock \showarticletitle{Gaussianhair: Hair modeling and rendering with light-aware gaussians}.
\newblock \bibinfo{journal}{\emph{arXiv preprint arXiv:2402.10483}} (\bibinfo{year}{2024}).
\newblock


\bibitem[Mao et~al\mbox{.}(2004)]%
        {mao2004sketch}
\bibfield{author}{\bibinfo{person}{Xiaoyang Mao}, \bibinfo{person}{Hiroki Kato}, \bibinfo{person}{Atsumi Imamiya}, {and} \bibinfo{person}{Ken Anjyo}.} \bibinfo{year}{2004}\natexlab{}.
\newblock \showarticletitle{Sketch interface based expressive hairstyle modelling and rendering}.
\newblock  (\bibinfo{date}{Jun} \bibinfo{year}{2004}).
\newblock


\bibitem[McCann and Pollard(2009)]%
        {mccan2009local}
\bibfield{author}{\bibinfo{person}{James McCann} {and} \bibinfo{person}{Nancy Pollard}.} \bibinfo{year}{2009}\natexlab{}.
\newblock \showarticletitle{Local layering}.
\newblock \bibinfo{journal}{\emph{ACM Transactions on Graphics}} \bibinfo{volume}{28}, \bibinfo{number}{3} (\bibinfo{date}{Jul} \bibinfo{year}{2009}), \bibinfo{pages}{1–7}.
\newblock
\href{https://doi.org/10.1145/1531326.1531390}{doi:\nolinkurl{10.1145/1531326.1531390}}


\bibitem[Nam et~al\mbox{.}(2019)]%
        {nam2019strand}
\bibfield{author}{\bibinfo{person}{Giljoo Nam}, \bibinfo{person}{Chenglei Wu}, \bibinfo{person}{Min~H Kim}, {and} \bibinfo{person}{Yaser Sheikh}.} \bibinfo{year}{2019}\natexlab{}.
\newblock \showarticletitle{Strand-accurate multi-view hair capture}. In \bibinfo{booktitle}{\emph{Proceedings of the IEEE/CVF Conference on Computer Vision and Pattern Recognition}}. \bibinfo{pages}{155--164}.
\newblock


\bibitem[Noble and Tang(2004)]%
        {paul2006model}
\bibfield{author}{\bibinfo{person}{Paul Noble} {and} \bibinfo{person}{Wen Tang}.} \bibinfo{year}{2004}\natexlab{}.
\newblock \showarticletitle{Modelling and Animating Cartoon Hair with NURBS Surfaces}. In \bibinfo{booktitle}{\emph{Proceedings of the Computer Graphics International}} \emph{(\bibinfo{series}{CGI '04})}. \bibinfo{publisher}{IEEE Computer Society}, \bibinfo{address}{USA}, \bibinfo{pages}{60–67}.
\newblock
\showISBNx{0769521711}


\bibitem[Paschalidou et~al\mbox{.}(2021)]%
        {paschalidou2021atiss}
\bibfield{author}{\bibinfo{person}{Despoina Paschalidou}, \bibinfo{person}{Amlan Kar}, \bibinfo{person}{Maria Shugrina}, \bibinfo{person}{Karsten Kreis}, \bibinfo{person}{Andreas Geiger}, {and} \bibinfo{person}{Sanja Fidler}.} \bibinfo{year}{2021}\natexlab{}.
\newblock \showarticletitle{Atiss: Autoregressive transformers for indoor scene synthesis}.
\newblock \bibinfo{journal}{\emph{Advances in Neural Information Processing Systems}}  \bibinfo{volume}{34} (\bibinfo{year}{2021}), \bibinfo{pages}{12013--12026}.
\newblock


\bibitem[Petrov et~al\mbox{.}(2024)]%
        {petrov2024gem3d}
\bibfield{author}{\bibinfo{person}{Dmitry Petrov}, \bibinfo{person}{Pradyumn Goyal}, \bibinfo{person}{Vikas Thamizharasan}, \bibinfo{person}{Vladimir Kim}, \bibinfo{person}{Matheus Gadelha}, \bibinfo{person}{Melinos Averkiou}, \bibinfo{person}{Siddhartha Chaudhuri}, {and} \bibinfo{person}{Evangelos Kalogerakis}.} \bibinfo{year}{2024}\natexlab{}.
\newblock \showarticletitle{Gem3d: Generative medial abstractions for 3d shape synthesis}. In \bibinfo{booktitle}{\emph{ACM SIGGRAPH 2024 Conference Papers}}. \bibinfo{pages}{1--11}.
\newblock


\bibitem[Piuze et~al\mbox{.}(2011)]%
        {piuze2011generalized}
\bibfield{author}{\bibinfo{person}{Emmanuel Piuze}, \bibinfo{person}{Paul~G Kry}, {and} \bibinfo{person}{Kaleem Siddiqi}.} \bibinfo{year}{2011}\natexlab{}.
\newblock \showarticletitle{Generalized helicoids for modeling hair geometry}. In \bibinfo{booktitle}{\emph{Computer Graphics Forum}}, Vol.~\bibinfo{volume}{30}. Wiley Online Library, \bibinfo{pages}{247--256}.
\newblock


\bibitem[Podell et~al\mbox{.}(2023)]%
        {podell2023sdxl}
\bibfield{author}{\bibinfo{person}{Dustin Podell}, \bibinfo{person}{Zion English}, \bibinfo{person}{Kyle Lacey}, \bibinfo{person}{Andreas Blattmann}, \bibinfo{person}{Tim Dockhorn}, \bibinfo{person}{Jonas M{\"u}ller}, \bibinfo{person}{Joe Penna}, {and} \bibinfo{person}{Robin Rombach}.} \bibinfo{year}{2023}\natexlab{}.
\newblock \showarticletitle{Sdxl: Improving latent diffusion models for high-resolution image synthesis}.
\newblock \bibinfo{journal}{\emph{arXiv preprint arXiv:2307.01952}} (\bibinfo{year}{2023}).
\newblock


\bibitem[Radford et~al\mbox{.}(2021)]%
        {radford2021learning}
\bibfield{author}{\bibinfo{person}{Alec Radford}, \bibinfo{person}{Jong~Wook Kim}, \bibinfo{person}{Chris Hallacy}, \bibinfo{person}{Aditya Ramesh}, \bibinfo{person}{Gabriel Goh}, \bibinfo{person}{Sandhini Agarwal}, \bibinfo{person}{Girish Sastry}, \bibinfo{person}{Amanda Askell}, \bibinfo{person}{Pamela Mishkin}, \bibinfo{person}{Jack Clark}, {et~al\mbox{.}}} \bibinfo{year}{2021}\natexlab{}.
\newblock \showarticletitle{Learning transferable visual models from natural language supervision}. In \bibinfo{booktitle}{\emph{International conference on machine learning}}. PmLR, \bibinfo{pages}{8748--8763}.
\newblock


\bibitem[Radford et~al\mbox{.}(2019)]%
        {radford2019language}
\bibfield{author}{\bibinfo{person}{Alec Radford}, \bibinfo{person}{Jeffrey Wu}, \bibinfo{person}{Rewon Child}, \bibinfo{person}{David Luan}, \bibinfo{person}{Dario Amodei}, \bibinfo{person}{Ilya Sutskever}, {et~al\mbox{.}}} \bibinfo{year}{2019}\natexlab{}.
\newblock \showarticletitle{Language models are unsupervised multitask learners}.
\newblock \bibinfo{journal}{\emph{OpenAI blog}} \bibinfo{volume}{1}, \bibinfo{number}{8} (\bibinfo{year}{2019}), \bibinfo{pages}{9}.
\newblock


\bibitem[Ritchie et~al\mbox{.}(2019)]%
        {ritchie2019fast}
\bibfield{author}{\bibinfo{person}{Daniel Ritchie}, \bibinfo{person}{Kai Wang}, {and} \bibinfo{person}{Yu-an Lin}.} \bibinfo{year}{2019}\natexlab{}.
\newblock \showarticletitle{Fast and flexible indoor scene synthesis via deep convolutional generative models}. In \bibinfo{booktitle}{\emph{Proceedings of the IEEE/CVF Conference on Computer Vision and Pattern Recognition}}. \bibinfo{pages}{6182--6190}.
\newblock


\bibitem[Rivers et~al\mbox{.}(2010)]%
        {river2010cartoon}
\bibfield{author}{\bibinfo{person}{Alec Rivers}, \bibinfo{person}{Takeo Igarashi}, {and} \bibinfo{person}{Frédo Durand}.} \bibinfo{year}{2010}\natexlab{}.
\newblock \showarticletitle{2.5D cartoon models}.
\newblock \bibinfo{journal}{\emph{ACM Transactions on Graphics}} \bibinfo{volume}{29}, \bibinfo{number}{4} (\bibinfo{date}{Jul} \bibinfo{year}{2010}), \bibinfo{pages}{1–7}.
\newblock
\href{https://doi.org/10.1145/1778765.1778796}{doi:\nolinkurl{10.1145/1778765.1778796}}


\bibitem[Rosu et~al\mbox{.}(2022)]%
        {rosu2022neuralstrands}
\bibfield{author}{\bibinfo{person}{Radu~Alexandru Rosu}, \bibinfo{person}{Shunsuke Saito}, \bibinfo{person}{Ziyan Wang}, \bibinfo{person}{Chenglei Wu}, \bibinfo{person}{Sven Behnke}, {and} \bibinfo{person}{Giljoo Nam}.} \bibinfo{year}{2022}\natexlab{}.
\newblock \showarticletitle{Neural strands: Learning hair geometry and appearance from multi-view images}. In \bibinfo{booktitle}{\emph{European Conference on Computer Vision}}. Springer, \bibinfo{pages}{73--89}.
\newblock


\bibitem[Rosu et~al\mbox{.}(2025)]%
        {rosu2025difflocks}
\bibfield{author}{\bibinfo{person}{Radu~Alexandru Rosu}, \bibinfo{person}{Keyu Wu}, \bibinfo{person}{Yao Feng}, \bibinfo{person}{Youyi Zheng}, {and} \bibinfo{person}{Michael~J Black}.} \bibinfo{year}{2025}\natexlab{}.
\newblock \showarticletitle{DiffLocks: Generating 3D Hair from a Single Image using Diffusion Models}. In \bibinfo{booktitle}{\emph{Proceedings of the Computer Vision and Pattern Recognition Conference}}. \bibinfo{pages}{10847--10857}.
\newblock


\bibitem[Saito et~al\mbox{.}(2018)]%
        {saito2018hair}
\bibfield{author}{\bibinfo{person}{Shunsuke Saito}, \bibinfo{person}{Liwen Hu}, \bibinfo{person}{Chongyang Ma}, \bibinfo{person}{Hikaru Ibayashi}, \bibinfo{person}{Linjie Luo}, {and} \bibinfo{person}{Hao Li}.} \bibinfo{year}{2018}\natexlab{}.
\newblock \showarticletitle{3D hair synthesis using volumetric variational autoencoders}.
\newblock \bibinfo{journal}{\emph{ACM Transactions on Graphics}} (\bibinfo{date}{Dec} \bibinfo{year}{2018}), \bibinfo{pages}{1–12}.
\newblock
\href{https://doi.org/10.1145/3272127.3275019}{doi:\nolinkurl{10.1145/3272127.3275019}}


\bibitem[Sakai and Savchenko(2013)]%
        {sakai2013skeleton}
\bibfield{author}{\bibinfo{person}{Takeyuki Sakai} {and} \bibinfo{person}{Vladimir Savchenko}.} \bibinfo{year}{2013}\natexlab{}.
\newblock \showarticletitle{Skeleton-based anime hair modeling and visualization}. In \bibinfo{booktitle}{\emph{2013 International Conference on Cyberworlds}}. IEEE, \bibinfo{pages}{318--321}.
\newblock


\bibitem[Shen et~al\mbox{.}(2023)]%
        {shen2023CT2Hair}
\bibfield{author}{\bibinfo{person}{Yuefan Shen}, \bibinfo{person}{Shunsuke Saito}, \bibinfo{person}{Ziyan Wang}, \bibinfo{person}{Olivier Maury}, \bibinfo{person}{Chenglei Wu}, \bibinfo{person}{Jessica Hodgins}, \bibinfo{person}{Youyi Zheng}, {and} \bibinfo{person}{Giljoo Nam}.} \bibinfo{year}{2023}\natexlab{}.
\newblock \showarticletitle{CT2Hair: High-Fidelity 3D Hair Modeling using Computed Tomography}.
\newblock \bibinfo{journal}{\emph{ACM Transactions on Graphics}} \bibinfo{volume}{42}, \bibinfo{number}{4}, Article \bibinfo{articleno}{75} (\bibinfo{year}{2023}), \bibinfo{numpages}{13}~pages.
\newblock


\bibitem[Shin et~al\mbox{.}(2006)]%
        {shin2006style}
\bibfield{author}{\bibinfo{person}{Jung Shin}, \bibinfo{person}{Michael Haller}, {and} \bibinfo{person}{R. Mukundan}.} \bibinfo{year}{2006}\natexlab{}.
\newblock \showarticletitle{A stylized cartoon hair renderer}. In \bibinfo{booktitle}{\emph{Proceedings of the 2006 ACM SIGCHI international conference on Advances in computer entertainment technology}}. \bibinfo{pages}{64}.
\newblock
\href{https://doi.org/10.1145/1178823.1178899}{doi:\nolinkurl{10.1145/1178823.1178899}}


\bibitem[Siddiqui et~al\mbox{.}(2024)]%
        {siddiqui2024meshgpt}
\bibfield{author}{\bibinfo{person}{Yawar Siddiqui}, \bibinfo{person}{Antonio Alliegro}, \bibinfo{person}{Alexey Artemov}, \bibinfo{person}{Tatiana Tommasi}, \bibinfo{person}{Daniele Sirigatti}, \bibinfo{person}{Vladislav Rosov}, \bibinfo{person}{Angela Dai}, {and} \bibinfo{person}{Matthias Nie{\ss}ner}.} \bibinfo{year}{2024}\natexlab{}.
\newblock \showarticletitle{Meshgpt: Generating triangle meshes with decoder-only transformers}. In \bibinfo{booktitle}{\emph{Proceedings of the IEEE/CVF Conference on Computer Vision and Pattern Recognition}}. \bibinfo{pages}{19615--19625}.
\newblock


\bibitem[Sklyarova et~al\mbox{.}(2023a)]%
        {sklyarova2023neural}
\bibfield{author}{\bibinfo{person}{Vanessa Sklyarova}, \bibinfo{person}{Jenya Chelishev}, \bibinfo{person}{Andreea Dogaru}, \bibinfo{person}{Igor Medvedev}, \bibinfo{person}{Victor Lempitsky}, {and} \bibinfo{person}{Egor Zakharov}.} \bibinfo{year}{2023}\natexlab{a}.
\newblock \showarticletitle{Neural haircut: Prior-guided strand-based hair reconstruction}. In \bibinfo{booktitle}{\emph{Proceedings of the IEEE/CVF International Conference on Computer Vision}}. \bibinfo{pages}{19762--19773}.
\newblock


\bibitem[Sklyarova et~al\mbox{.}(2023b)]%
        {sklyarova2023haar}
\bibfield{author}{\bibinfo{person}{Vanessa Sklyarova}, \bibinfo{person}{Egor Zakharov}, \bibinfo{person}{Otmar Hilliges}, \bibinfo{person}{Michael~J Black}, {and} \bibinfo{person}{Justus Thies}.} \bibinfo{year}{2023}\natexlab{b}.
\newblock \showarticletitle{Haar: Text-conditioned generative model of 3d strand-based human hairstyles}.
\newblock \bibinfo{journal}{\emph{arXiv preprint arXiv:2312.11666}} (\bibinfo{year}{2023}).
\newblock


\bibitem[Takimoto et~al\mbox{.}(2024)]%
        {takimoto2024dr}
\bibfield{author}{\bibinfo{person}{Yusuke Takimoto}, \bibinfo{person}{Hikari Takehara}, \bibinfo{person}{Hiroyuki Sato}, \bibinfo{person}{Zihao Zhu}, {and} \bibinfo{person}{Bo Zheng}.} \bibinfo{year}{2024}\natexlab{}.
\newblock \showarticletitle{Dr. Hair: Reconstructing Scalp-Connected Hair Strands without Pre-Training via Differentiable Rendering of Line Segments}. In \bibinfo{booktitle}{\emph{Proceedings of the IEEE/CVF Conference on Computer Vision and Pattern Recognition}}. \bibinfo{pages}{20601--20611}.
\newblock


\bibitem[VRoid(2022)]%
        {vroidHub}
\bibfield{author}{\bibinfo{person}{VRoid}.} \bibinfo{year}{2022}\natexlab{}.
\newblock \bibinfo{title}{VRoid Hub}.
\newblock
\urldef\tempurl%
\url{https://vroid.com/}
\showURL{%
\tempurl}


\bibitem[Wang et~al\mbox{.}(2024b)]%
        {wang2024nova}
\bibfield{author}{\bibinfo{person}{Hongsheng Wang}, \bibinfo{person}{Xinrui Zhou}, {and} \bibinfo{person}{Feng Lin}.} \bibinfo{year}{2024}\natexlab{b}.
\newblock \showarticletitle{NOVA-3D: Non-overlapped Views for 3D Anime Character Reconstruction}. In \bibinfo{booktitle}{\emph{Proceedings of the 6th ACM International Conference on Multimedia in Asia Workshops}}. \bibinfo{pages}{1--7}.
\newblock


\bibitem[Wang et~al\mbox{.}(2021)]%
        {wang2021sceneformer}
\bibfield{author}{\bibinfo{person}{Xinpeng Wang}, \bibinfo{person}{Chandan Yeshwanth}, {and} \bibinfo{person}{Matthias Nie{\ss}ner}.} \bibinfo{year}{2021}\natexlab{}.
\newblock \showarticletitle{Sceneformer: Indoor scene generation with transformers}. In \bibinfo{booktitle}{\emph{2021 International Conference on 3D Vision (3DV)}}. IEEE, \bibinfo{pages}{106--115}.
\newblock


\bibitem[Wang et~al\mbox{.}(2024a)]%
        {wang2024llamameshunifying3dmesh}
\bibfield{author}{\bibinfo{person}{Zhengyi Wang}, \bibinfo{person}{Jonathan Lorraine}, \bibinfo{person}{Yikai Wang}, \bibinfo{person}{Hang Su}, \bibinfo{person}{Jun Zhu}, \bibinfo{person}{Sanja Fidler}, {and} \bibinfo{person}{Xiaohui Zeng}.} \bibinfo{year}{2024}\natexlab{a}.
\newblock \bibinfo{title}{LLaMA-Mesh: Unifying 3D Mesh Generation with Language Models}.
\newblock
\showeprint[arxiv]{2411.09595}~[cs.LG]
\urldef\tempurl%
\url{https://arxiv.org/abs/2411.09595}
\showURL{%
\tempurl}


\bibitem[Weng et~al\mbox{.}(2024)]%
        {weng2024scaling}
\bibfield{author}{\bibinfo{person}{Haohan Weng}, \bibinfo{person}{Zibo Zhao}, \bibinfo{person}{Biwen Lei}, \bibinfo{person}{Xianghui Yang}, \bibinfo{person}{Jian Liu}, \bibinfo{person}{Zeqiang Lai}, \bibinfo{person}{Zhuo Chen}, \bibinfo{person}{Yuhong Liu}, \bibinfo{person}{Jie Jiang}, \bibinfo{person}{Chunchao Guo}, \bibinfo{person}{Tong Zhang}, \bibinfo{person}{Shenghua Gao}, {and} \bibinfo{person}{C.~L.~Philip Chen}.} \bibinfo{year}{2024}\natexlab{}.
\newblock \showarticletitle{Scaling Mesh Generation via Compressive Tokenization}.
\newblock \bibinfo{journal}{\emph{arXiv preprint arXiv:2411.07025}} (\bibinfo{year}{2024}).
\newblock


\bibitem[Wu et~al\mbox{.}(2024)]%
        {wu2024monohair}
\bibfield{author}{\bibinfo{person}{Keyu Wu}, \bibinfo{person}{Lingchen Yang}, \bibinfo{person}{Zhiyi Kuang}, \bibinfo{person}{Yao Feng}, \bibinfo{person}{Xutao Han}, \bibinfo{person}{Yuefan Shen}, \bibinfo{person}{Hongbo Fu}, \bibinfo{person}{Kun Zhou}, {and} \bibinfo{person}{Youyi Zheng}.} \bibinfo{year}{2024}\natexlab{}.
\newblock \showarticletitle{Monohair: High-fidelity hair modeling from a monocular video}. In \bibinfo{booktitle}{\emph{Proceedings of the IEEE/CVF Conference on Computer Vision and Pattern Recognition}}. \bibinfo{pages}{24164--24173}.
\newblock


\bibitem[Xu and Yang(2001)]%
        {xu2001vhair}
\bibfield{author}{\bibinfo{person}{Zhan Xu} {and} \bibinfo{person}{Xue~Dong Yang}.} \bibinfo{year}{2001}\natexlab{}.
\newblock \showarticletitle{V-HairStudio: an interactive tool for hair design}.
\newblock \bibinfo{journal}{\emph{IEEE Computer Graphics and Applications}} \bibinfo{volume}{21}, \bibinfo{number}{1} (\bibinfo{date}{Jan} \bibinfo{year}{2001}), \bibinfo{pages}{36–43}.
\newblock
\href{https://doi.org/10.1109/38.920625}{doi:\nolinkurl{10.1109/38.920625}}


\bibitem[Yan et~al\mbox{.}(2024)]%
        {yan2024frankenstein}
\bibfield{author}{\bibinfo{person}{Han Yan}, \bibinfo{person}{Yang Li}, \bibinfo{person}{Zhennan Wu}, \bibinfo{person}{Shenzhou Chen}, \bibinfo{person}{Weixuan Sun}, \bibinfo{person}{Taizhang Shang}, \bibinfo{person}{Weizhe Liu}, \bibinfo{person}{Tian Chen}, \bibinfo{person}{Xiaqiang Dai}, \bibinfo{person}{Chao Ma}, {et~al\mbox{.}}} \bibinfo{year}{2024}\natexlab{}.
\newblock \showarticletitle{Frankenstein: Generating semantic-compositional 3d scenes in one tri-plane}. In \bibinfo{booktitle}{\emph{SIGGRAPH Asia 2024 Conference Papers}}. \bibinfo{pages}{1--11}.
\newblock


\bibitem[Yang et~al\mbox{.}(2000)]%
        {yang2000the}
\bibfield{author}{\bibinfo{person}{Xue~Dong Yang}, \bibinfo{person}{Zhan Xu}, \bibinfo{person}{Jun Yang}, {and} \bibinfo{person}{Tao Wang}.} \bibinfo{year}{2000}\natexlab{}.
\newblock \showarticletitle{The Cluster Hair Model}.
\newblock \bibinfo{journal}{\emph{Graphical Models}} (\bibinfo{date}{Mar} \bibinfo{year}{2000}), \bibinfo{pages}{85–103}.
\newblock
\href{https://doi.org/10.1006/gmod.1999.0518}{doi:\nolinkurl{10.1006/gmod.1999.0518}}


\bibitem[Ye et~al\mbox{.}(2025)]%
        {ye2025primitiveanything}
\bibfield{author}{\bibinfo{person}{Jingwen Ye}, \bibinfo{person}{Yuze He}, \bibinfo{person}{Yanning Zhou}, \bibinfo{person}{Yiqin Zhu}, \bibinfo{person}{Kaiwen Xiao}, \bibinfo{person}{Yong-Jin Liu}, \bibinfo{person}{Wei Yang}, {and} \bibinfo{person}{Xiao Han}.} \bibinfo{year}{2025}\natexlab{}.
\newblock \bibinfo{title}{PrimitiveAnything: Human-Crafted 3D Primitive Assembly Generation with Auto-Regressive Transformer}.
\newblock
\showeprint[arxiv]{2505.04622}~[cs.GR]


\bibitem[Yeh et~al\mbox{.}(2014)]%
        {yeh20142}
\bibfield{author}{\bibinfo{person}{Chih-Kuo Yeh}, \bibinfo{person}{Pradeep~Kumar Jayaraman}, \bibinfo{person}{Xiaopei Liu}, \bibinfo{person}{Chi-Wing Fu}, {and} \bibinfo{person}{Tong-Yee Lee}.} \bibinfo{year}{2014}\natexlab{}.
\newblock \showarticletitle{2.5 D cartoon hair modeling and manipulation}.
\newblock \bibinfo{journal}{\emph{IEEE Transactions on Visualization and computer graphics}} \bibinfo{volume}{21}, \bibinfo{number}{3} (\bibinfo{year}{2014}), \bibinfo{pages}{304--314}.
\newblock


\bibitem[Yuksel et~al\mbox{.}(2009)]%
        {yuksel2009hairmeshes}
\bibfield{author}{\bibinfo{person}{Cem Yuksel}, \bibinfo{person}{Scott Schaefer}, {and} \bibinfo{person}{John Keyser}.} \bibinfo{year}{2009}\natexlab{}.
\newblock \showarticletitle{Hair meshes}.
\newblock \bibinfo{journal}{\emph{ACM Transactions on Graphics (TOG)}} \bibinfo{volume}{28}, \bibinfo{number}{5} (\bibinfo{year}{2009}), \bibinfo{pages}{1--7}.
\newblock


\bibitem[Zakharov et~al\mbox{.}(2024)]%
        {zakharov2024human}
\bibfield{author}{\bibinfo{person}{Egor Zakharov}, \bibinfo{person}{Vanessa Sklyarova}, \bibinfo{person}{Michael Black}, \bibinfo{person}{Giljoo Nam}, \bibinfo{person}{Justus Thies}, {and} \bibinfo{person}{Otmar Hilliges}.} \bibinfo{year}{2024}\natexlab{}.
\newblock \showarticletitle{Human hair reconstruction with strand-aligned 3d gaussians}. In \bibinfo{booktitle}{\emph{European Conference on Computer Vision}}. Springer, \bibinfo{pages}{409--425}.
\newblock


\bibitem[Zhang et~al\mbox{.}(2012)]%
        {zhang2012excol}
\bibfield{author}{\bibinfo{person}{Lei Zhang}, \bibinfo{person}{Hua Huang}, {and} \bibinfo{person}{Hongbo Fu}.} \bibinfo{year}{2012}\natexlab{}.
\newblock \showarticletitle{EXCOL: An EXtract-and-COmplete Layering Approach to Cartoon Animation Reusing}.
\newblock \bibinfo{journal}{\emph{IEEE Transactions on Visualization and Computer Graphics,IEEE Transactions on Visualization and Computer Graphics}} (\bibinfo{date}{Jul} \bibinfo{year}{2012}).
\newblock


\bibitem[Zhang et~al\mbox{.}(2024)]%
        {zhang2024clay}
\bibfield{author}{\bibinfo{person}{Longwen Zhang}, \bibinfo{person}{Ziyu Wang}, \bibinfo{person}{Qixuan Zhang}, \bibinfo{person}{Qiwei Qiu}, \bibinfo{person}{Anqi Pang}, \bibinfo{person}{Haoran Jiang}, \bibinfo{person}{Wei Yang}, \bibinfo{person}{Lan Xu}, {and} \bibinfo{person}{Jingyi Yu}.} \bibinfo{year}{2024}\natexlab{}.
\newblock \showarticletitle{CLAY: A Controllable Large-scale Generative Model for Creating High-quality 3D Assets}.
\newblock \bibinfo{journal}{\emph{ACM Transactions on Graphics (TOG)}} \bibinfo{volume}{43}, \bibinfo{number}{4} (\bibinfo{year}{2024}), \bibinfo{pages}{1--20}.
\newblock


\bibitem[Zhang et~al\mbox{.}(2020)]%
        {zhang2020h3dnet}
\bibfield{author}{\bibinfo{person}{Zaiwei Zhang}, \bibinfo{person}{Bo Sun}, \bibinfo{person}{Haitao Yang}, {and} \bibinfo{person}{Qixing Huang}.} \bibinfo{year}{2020}\natexlab{}.
\newblock \showarticletitle{H3dnet: 3d object detection using hybrid geometric primitives}. In \bibinfo{booktitle}{\emph{Computer Vision--ECCV 2020: 16th European Conference, Glasgow, UK, August 23--28, 2020, Proceedings, Part XII 16}}. Springer, \bibinfo{pages}{311--329}.
\newblock


\bibitem[Zhao et~al\mbox{.}(2025)]%
        {zhao2025deepmesh}
\bibfield{author}{\bibinfo{person}{Ruowen Zhao}, \bibinfo{person}{Junliang Ye}, \bibinfo{person}{Zhengyi Wang}, \bibinfo{person}{Guangce Liu}, \bibinfo{person}{Yiwen Chen}, \bibinfo{person}{Yikai Wang}, {and} \bibinfo{person}{Jun Zhu}.} \bibinfo{year}{2025}\natexlab{}.
\newblock \showarticletitle{DeepMesh: Auto-Regressive Artist-mesh Creation with Reinforcement Learning}.
\newblock \bibinfo{journal}{\emph{arXiv preprint arXiv:2503.15265}} (\bibinfo{year}{2025}).
\newblock


\bibitem[Zhao et~al\mbox{.}(2023)]%
        {zhao2023michelangelo}
\bibfield{author}{\bibinfo{person}{Zibo Zhao}, \bibinfo{person}{Wen Liu}, \bibinfo{person}{Xin Chen}, \bibinfo{person}{Xianfang Zeng}, \bibinfo{person}{Rui Wang}, \bibinfo{person}{Pei Cheng}, \bibinfo{person}{BIN FU}, \bibinfo{person}{Tao Chen}, \bibinfo{person}{Gang YU}, {and} \bibinfo{person}{Shenghua Gao}.} \bibinfo{year}{2023}\natexlab{}.
\newblock \showarticletitle{Michelangelo: Conditional 3D Shape Generation based on Shape-Image-Text Aligned Latent Representation}. In \bibinfo{booktitle}{\emph{Thirty-seventh Conference on Neural Information Processing Systems}}.
\newblock
\urldef\tempurl%
\url{https://openreview.net/forum?id=xmxgMij3LY}
\showURL{%
\tempurl}


\bibitem[Zheng et~al\mbox{.}(2023)]%
        {zheng2023hairstep}
\bibfield{author}{\bibinfo{person}{Yujian Zheng}, \bibinfo{person}{Zirong Jin}, \bibinfo{person}{Moran Li}, \bibinfo{person}{Haibin Huang}, \bibinfo{person}{Chongyang Ma}, \bibinfo{person}{Shuguang Cui}, {and} \bibinfo{person}{Xiaoguang Han}.} \bibinfo{year}{2023}\natexlab{}.
\newblock \showarticletitle{Hairstep: Transfer synthetic to real using strand and depth maps for single-view 3d hair modeling}. In \bibinfo{booktitle}{\emph{Proceedings of the IEEE/CVF Conference on Computer Vision and Pattern Recognition}}. \bibinfo{pages}{12726--12735}.
\newblock


\bibitem[Zhou et~al\mbox{.}(2023)]%
        {zhou2023groomgen}
\bibfield{author}{\bibinfo{person}{Yuxiao Zhou}, \bibinfo{person}{Menglei Chai}, \bibinfo{person}{Alessandro Pepe}, \bibinfo{person}{Markus Gross}, {and} \bibinfo{person}{Thabo Beeler}.} \bibinfo{year}{2023}\natexlab{}.
\newblock \showarticletitle{Groomgen: A high-quality generative hair model using hierarchical latent representations}.
\newblock \bibinfo{journal}{\emph{ACM Transactions on Graphics (TOG)}} \bibinfo{volume}{42}, \bibinfo{number}{6} (\bibinfo{year}{2023}), \bibinfo{pages}{1--16}.
\newblock


\end{thebibliography}

\clearpage
\appendix
\setcounter{page}{1}

\section{Extra Inference Strategies}
Given the inherent priors in hairstyle structure, we have developed specialized inference strategies to improve generation quality. One critical challenge is that incorrectly positioned hair roots can propagate errors throughout the entire strand and subsequently affect the generation of subsequent hairs. To address this issue, we implement a \textit{Root Position Verification} procedure that examines the distance between generated root positions and the reference point cloud. When this distance exceeds a predetermined threshold (set at 0.03 in our implementation), we identify the root as potentially erroneous.
In such cases, we examine the top 10 most probable alternative xyz position combinations predicted by our model, ranked by probability. We sequentially test these alternatives until we find a position that falls within the threshold distance. If no suitable position is found among these candidates, we select the position combination with the minimum distance to the point cloud. After establishing a corrected root position, we resume auto-regressive generation from this point.

Another challenge involves cases where excessively long hair strands are generated, which can interfere with the model's understanding of the overall hairstyle structure. To mitigate this issue, we employ a {\it Length Normalization} strategy. Specifically, after generating each hair strand, if the number of control points exceeds 80, we interpolate the strand using a cubic spline to reduce it to 80 points. Furthermore, if a hair strand reaches 100 control points during generation, we immediately terminate it, interpolate to 80 points, and append an MOS token. These thresholds were established based on the statistical distribution of hair strand lengths in our dataset (as illustrated in Fig.~\ref{fig:data}).

\begin{table}[htbp]
\centering
\caption{Geometric comparison of different inference strategies.}
\resizebox{1.0\linewidth}{!}{
\begin{tabular}{l|c|c|c|c}
\toprule
Method & CD~$\downarrow$ & EMD~$\downarrow$ & Hausdorff~$\downarrow$ & Voxel-IoU~$\uparrow$ \\
\midrule
No extra strategy            & 0.0117     & 0.0128       & 0.0497       & 0.7566           \\
+~Root Position Verification  & 0.0117     & 0.0128       & 0.0481       & 0.7585           \\
+~Length Normalization & 0.0103 & 0.0117 & 0.0468 & 0.7676 \\
\bottomrule
\end{tabular}
}
\label{tab:infer}
\end{table}

Tab.~\ref{tab:infer} demonstrates the effectiveness of these inference strategies. The {\it Root Position Verification} shows improvements in Hausdorff distance and Voxel-IoU metrics, indicating a reduction in outlier points. The {\it Length Normalization} strategy yields significant improvements across all metrics, demonstrating the importance of maintaining appropriate control point lengths for generating coherent and geometrically accurate hairstyles.

\section{Details of Dataset}
\label{sec:dataset}
\xhdr{Comparison with other datasets}
As illustrated in Tab.~\ref{tab:datasets}, existing hair datasets primarily concentrate on realistic hairstyles, representing individual strands as polylines or curve segments devoid of three-dimensional volume.
In contrast, our AnimeHair dataset represents the \textbf{first} comprehensive collection specifically targeting anime-style hairstyles. With 37K hairstyles, it constitutes a large-scale resource that captures the rich diversity and stylistic variation found in anime character designs. Each hairstyle in our dataset preserves the full three-dimensional geometric information of individual hair elements, including their volumetric properties, spatial relationships, and distinctive structural patterns.

\begin{table}[h]
\centering
\caption{Comparison of current hair datasets.}
\label{tab:datasets}
\begin{tabular}{lcc}
\toprule
\textbf{Dataset} & \textbf{Total} & \textbf{Data Type} \\
\midrule
USC-HairSalon    & 343           & Realistic \\
Hair20k          & 3.7K         & Realistic \\
CT2Hair          & 10            & Realistic \\
GroomGen         & 7.7K         & Realistic \\
HAAR             & 9.8K         & Realistic \\
HiSa\&HiDa       & 1.3K         & Realistic \\
HairNet          & 40K       & Realistic \\
MultiHair        & 10K        & Realistic \\
AnimeHair~(Ours)  & 37K        & \textbf{Anime-style} \\
\bottomrule
\end{tabular}
\end{table}

\begin{figure*}[t]
\centering
\includegraphics[width=1\linewidth]{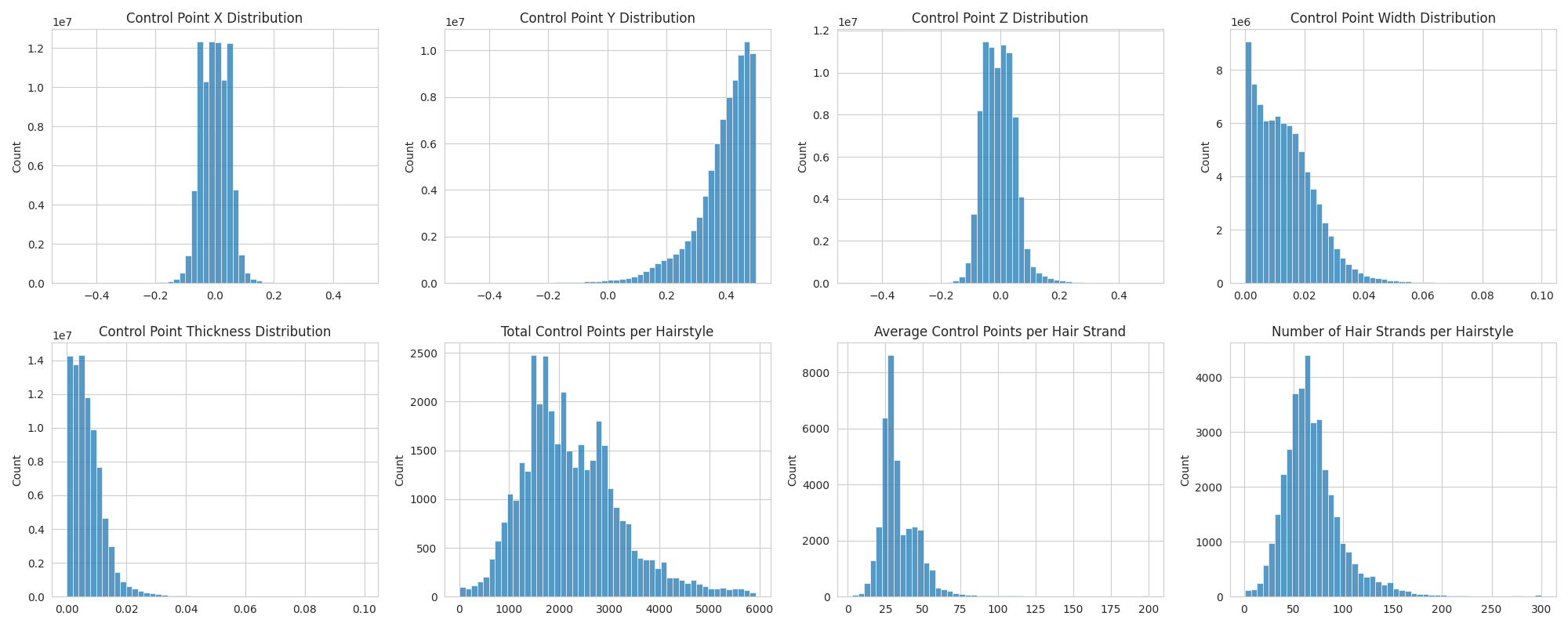}
\caption{Statistical analysis of AnimeHair data parameters.}
\label{fig:data}
\end{figure*}

\xhdr{Statistics}
To ensure consistent scale across all hairstyles in our dataset, we normalized each original 3D character model to the range $[-0.5, 0.5]$ and extracted the corresponding hair positions. Fig.~\ref{fig:data} illustrates the distribution characteristics of our dataset, including the spatial distribution of control point coordinates (x, y, z), width and thickness parameters, the number of hair per hairstyle, the average number of control points per hair, and the total number of control points per hairstyle.
The analysis reveals that most hairstyles contain between 25 and 130 individual hair strands, with each strand typically comprising 20 to 60 control points. The total number of control points per hairstyle generally falls within the range of 1,000 to 6,000, making them suitable for autoregressive generation approaches.

The distribution of hair card attributes, including xyz coordinates, width, and thickness, exhibits domain-specific characteristics that reflect the inherent properties of anime hairstyles. These attributes demonstrate broad value ranges with smooth distributions, indicating the diverse yet coherent nature of anime hair geometry. The distribution of control points per hair card, along with the total control point count, captures the structural complexity characteristic of anime-style hairstyles while maintaining computational tractability for transformer-based modeling.
Hair length analysis reveals that our dataset contains short hair (normalized length <0.25) in 51.4\% of cases, medium hair (0.25-0.40) in 33.0\%, and long hair (>0.40) in 15.6\% of instances.
Notably, anime hairstyles exhibit fundamental differences from realistic hair in several key aspects: they typically contain fewer total cards/strands, display significant inter-strand variations, and feature individual cards/strands that are neither purely straight nor curly but follow stylized geometric patterns. Each hairstyle demonstrates a high degree of artistic stylization that prioritizes visual appeal and character expression over physical realism (detailed analysis provided in Supplementary Sec.~D).

\xhdr{Preprocess}
Following the data collection process in PAniC-3D~\cite{chen2023panic} and StdGEN~\cite{he2024stdgen}, we gathered 3D anime characters from VRoid-Hub~\cite{vroidHub}, and implemented a custom filter that identifies hair elements based on model materials to extract hair mesh for subsequent processing.
The original hair mesh data presented several quality issues, including: absence of pre-separated individual hairs; non-watertight representations; fragmented single hair strands composed of multiple disconnected segments; and irregular mesh connections at endpoints with numerous combinatorial patterns, making the preprocessing of the original mesh challenging.
Our preprocessing pipeline began with the merging of extremely close vertices and isolating watertight components. We then identified irregular connection regions to locate endpoints and verified whether each hair strand, starting from these endpoints, conformed to the standard pattern of sequentially connected repeating units. After completing this verification for all components, we calculated the recall rate between the vertices that passed our quality checks and the original mesh vertices. Only hairstyles achieving a recall rate exceeding 98\% were incorporated into our dataset.
Additionally, we implemented filtering criteria to remove anomalous outliers, excluding cases with thickness or width values greater than 0.10, as well as hairstyles containing more than 6,000 total control points. This comprehensive preprocessing approach ensured that our dataset contained only structurally consistent hair representations suitable for training our generative model.

\section{Implementation Details}
\label{sec:implement}
Our transformer architecture consists of 6 layers with a hidden size of 768. The control point encoder is implemented as a single linear projection layer. The decoders are implemented as a 2-layer MLP with a hidden size of 768.
The smoothness weight $\lambda$ in Eq.~3 is set to 1.0.
The position attribute was discretized into 512 levels, while width and thickness were discretized into 128 levels. The embedding of each attribute was represented using 16-dimensional vectors. We set the maximum token length to 8,192.
For training, we used Adam Optimizer with a learning rate of 1e-4, using a batch size of 16 and accumulation of gradients in 4 steps. The model was trained on 8 V100 GPUs, and the entire training process took approximately 3 days to complete.

\xhdr{Shape Standardization}
In our current AnimeHair dataset, the repeating units are divided into two main types: those with diamond-shaped bases and those with triangular isosceles bases. Given the visual and geometric similarities between these two forms, we opted for standardization to enhance consistency and clarity of implementation. Specifically, during training, we transform all isosceles triangular bases into diamond shapes by reflecting them along their base edges.
This standardization approach has proven sufficient for producing hairstyles that meet the requirements of downstream applications such as video games and virtual reality environments. While our current implementation focuses on this unified representation, we acknowledge the potential for future work to explore generation methods that accommodate diverse repeating unit types.

\begin{figure*}[t]
\centering
\includegraphics[width=1.0\linewidth]{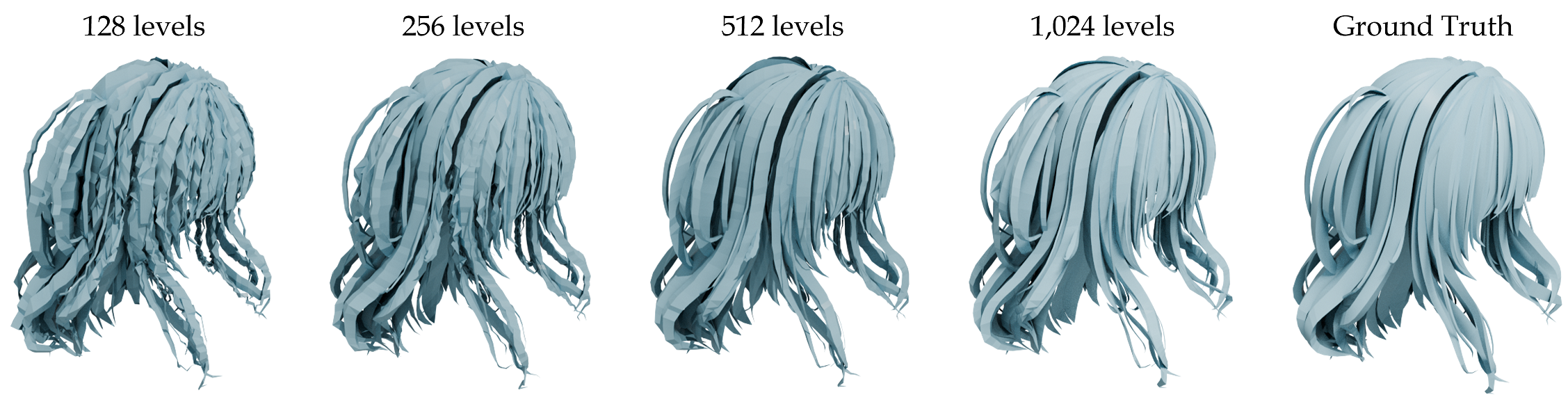}
\caption{Visualizations of applying different discretization levels.}
\label{fig:levels}
\end{figure*}

\xhdr{Piecewise Discretization}
Due to the highly non-uniform distribution of hairstyle control point attributes (as illustrated in Fig.~\ref{fig:data}), we implemented a piecewise discretization scheme to model these distributions more accurately and efficiently. The observed distributions show that position attributes (x, y, z) exhibit clustering patterns—x and z coordinates are more densely distributed near the origin, while y coordinates show higher density toward the top of the head. Similarly, width and thickness attributes display greater concentration at smaller values.
To address these distribution characteristics, we developed specific discretization rules for each attribute:

For x-coordinates, we established three intervals: $[-0.5, -0.1]$, $[-0.1, 0.1]$, and $[0.1, 0.5]$, with 96, 320, and 96 discrete levels respectively. This allocation dedicates higher resolution to the central region where most control points are concentrated.

For y-coordinates, we defined intervals at $[-0.5, 0]$, $[0, 0.3]$, and $[0.3, 0.5]$, allocating 96, 160, and 256 discrete levels respectively. This scheme provides increased resolution for the upper head region where hair detail and variation are most pronounced.

For z-coordinates, we established three intervals at $[-0.5, -0.15]$, $[-0.15, 0.1]$, and $[0.1, 0.5]$, with 96, 320, and 96 discrete levels respectively, following a similar pattern to the x-coordinate discretization.

Width attributes were divided into two intervals: $[0, 0.03]$ and $[0.03, 0.1]$, each allocated 64 discrete levels, while thickness attributes were similarly partitioned into intervals of $[0, 0.02]$ and $[0.02, 0.1]$, also with 64 discrete levels each.

This piecewise discretization approach enables a more precise representation of attribute distributions while maintaining computational efficiency.

\xhdr{Sampling Strategy of Point Clouds}
For the point cloud sampling strategy, we sample a fixed set of 10,000 points and their corresponding normals from the mesh surface. The performance could potentially be enhanced by increasing the number of sample points or by employing more advanced sampling distributions, such as Dora~\cite{chen2025dora}, which may provide a more beneficial point distribution.

\xhdr{Integration with StdGEN}
To integrate our framework with StdGEN~\cite{he2024stdgen}, we first construct a dataset from non-decomposed 3D hair meshes generated by the StdGEN pipeline. We render character images from six horizontal viewpoints (consistent with StdGEN's S-LRM input, at angles of 0, 45, 90, 180, 270, and 315 degrees relative to the front view). These renderings are then input to S-LRM to generate coarse 3D models, which are subsequently refined through multi-layer refinement to produce 3D hair meshes.
We sample point clouds from the surface of these meshes and combine them with ground-truth 3D hair mesh samples at a 50:50 ratio during training. Both types of point clouds are trained to predict our control-point-based 3D hairstyle representation.
During inference, we feed images of characters in arbitrary poses directly into the complete StdGEN pipeline, isolate the resulting hair mesh, sample surface point clouds, and process these through our CHARM pipeline. We then apply color back-projection using the same configuration as StdGEN to produce ready-to-use hairstyles that are compatible with the 3D character model.

\xhdr{Image-conditioned Generation}
For image-conditioned generation, we adopt a strategy similar to TANGLED~\cite{long2025tangled}. We construct an input image dataset by rendering character head photos, isolating the hair regions, and converting them to sketch format. These sketches are then processed through a DINO~\cite{caron2021emerging} encoder to generate image tokens. These tokens serve as conditions for auto-regressive generation through cross-attention mechanisms.
During inference, we employ the Segment Anything Model (SAM)~\cite{kirillov2023segany} to isolate the hair region from input images. The segmented hair is subsequently converted to a sketch format, which then serves as the conditioning input for the generation process.

\section{More Discussions}
\xhdr{Choice of Diffusion versus Auto-regressive Models}
Recent works in realistic hairstyle generation, such as HAAR~\cite{sklyarova2023haar} and TANGLED~\cite{long2025tangled}, employ diffusion models as their generative backbone. In contrast, our work adopts an auto-regressive approach. This methodological divergence stems from fundamental differences between anime-style and realistic hairstyles.

Anime hairstyles typically feature fewer individual hairs compared to realistic counterparts, yet each hair exhibits substantially higher variability in position, shape, and density. Additionally, anime hairstyles are characterized by non-fixed root positions and variable numbers of repeating units per individual hair. Realistic hair, conversely, typically maintains fixed root positions on the scalp, making it amenable to residual position prediction, planar unwrapping, and feature map modeling approaches. The high strand count and greater similarity between adjacent strands in realistic hair lend themselves to representation structures resembling images or feature maps, making diffusion models a natural fit. Anime hairstyles, characterized by a lower hair count but higher inter-hair variability, pose significant challenges for diffusion framework design. Their discrete and structured nature, combined with the frequent presence of substantial geometric and stylistic differences between adjacent hairs, makes them more well-suited for the sequential, element-by-element generation process of auto-regressive modeling.

\xhdr{Choice of discretization level}
The choice of discretization level requires careful consideration of both fidelity requirements and learning complexity. As illustrated in Fig.~\ref{fig:levels}, while a 128-level discretization may be sufficient for general mesh prediction~\cite{chen2024meshanything,wang2024llamameshunifying3dmesh} and primitive generation~\cite{ye2025primitiveanything} tasks, it produces notably inadequate results for hair reconstruction. At this lower discretization level, the generated hairstyles exhibit significant artifacts, including irregular structures and pixelation effects that compromise the smooth, flowing aesthetic essential to anime-style hair.
We selected a 512-level discretization scheme, which represents an optimal balance between representational capacity and computational feasibility.

\xhdr{Limitations and Future Works}
Our approach is subject to certain limitations arising from the inherent characteristics of our data sources. The AnimeHair dataset contains a higher proportion of feminine hairstyles compared to masculine styles, which may introduce representational biases in the generated outputs.
Future work could address this imbalance by developing a more demographically diverse training dataset with equitable representation across gender expressions, or by implementing specialized techniques to ensure fairness and inclusivity in the generated hairstyles.
Additionally, there are notable differences between anime-style and realistic-style data. Anime hairstyles lack certain morphological variations such as permed hair, upward-styled curls, and hair combed to both sides of the head, while realistic hairstyles lack distinctive anime features like frontal bangs (as shown in Sec.~4.5). These domain-specific gaps can lead to suboptimal cross-domain performance when methods trained on one domain are applied to the other. Future research could explore the development of cross-domain aligned datasets or domain adaptation techniques to enhance performance across realistic and anime input conditions and generation frameworks.

\begin{figure}[t]
\centering
\includegraphics[width=1\linewidth]{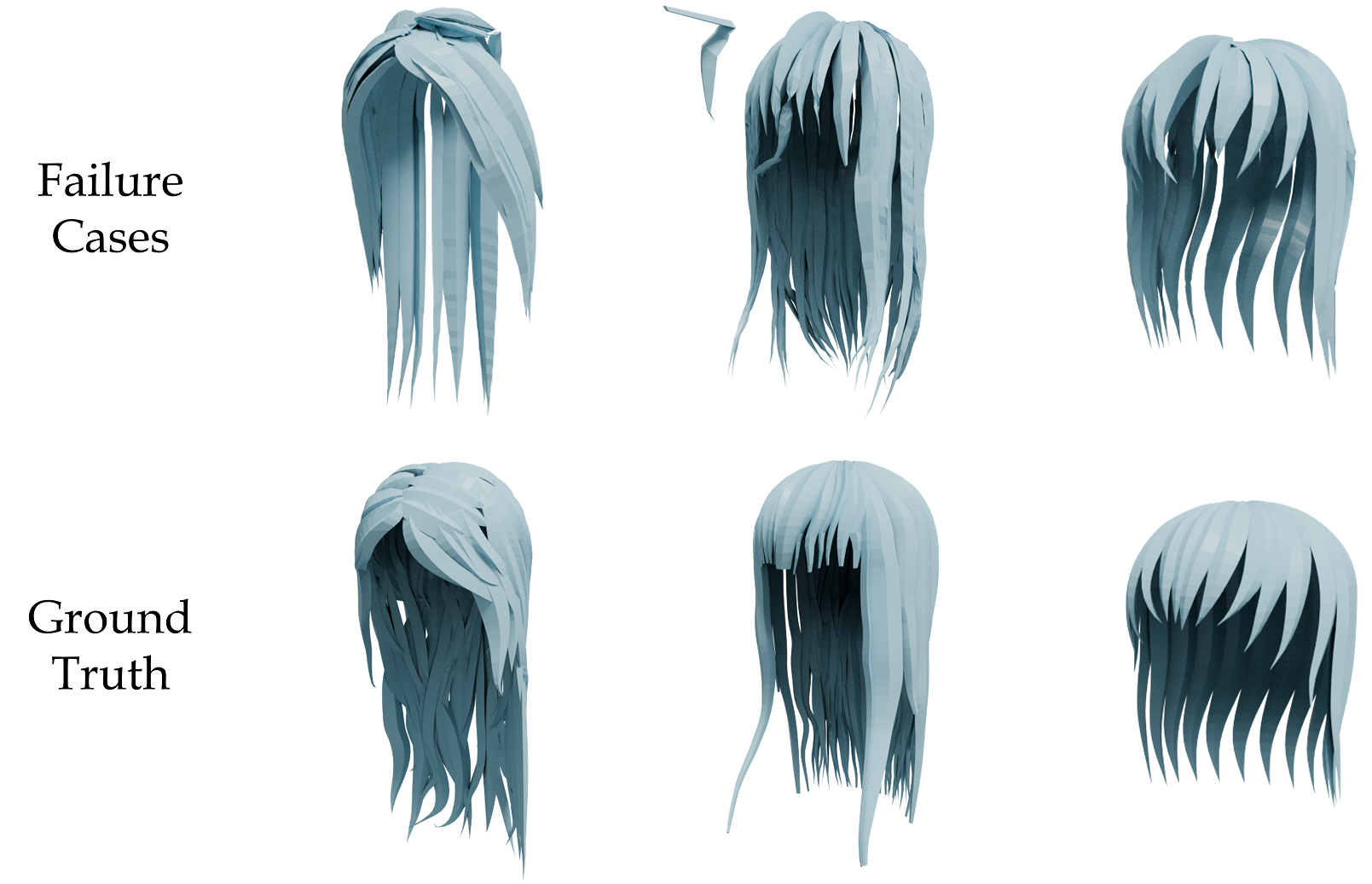}
\caption{Failure cases of our autoregressive generation approach. Left: incomplete hairstyle due to repetitive hair card generation at the same location. Middle: erroneous positional jump generating a hair card away from the main hairstyle. Right: low coverage result with hair gaps compared to ground truth.
}
\label{fig:fail}
\end{figure}

Our method inherits common limitations associated with autoregressive frameworks, including the occasional generation of repetitive control points and hair cards at identical locations, as well as erroneous positional jumps during the generation process. As illustrated in Fig.~\ref{fig:fail}, the leftmost case demonstrates how redundant hair card generation at the same position results in an incomplete hairstyle. The middle case exhibits an erroneous jump where a hair card is generated at a location distant from the main hairstyle region, though the model subsequently self-corrects this error. Furthermore, our dataset contains substantial variation in hair width and hair card count (as shown in Fig.~\ref{fig:data}), which can sometimes lead to generated output with lower coverage and hair gaps compared to the ground truth, as evidenced in the rightmost case. Future work could address these issues by incorporating hair-card-level constraints that penalize redundant generation, positional jumps, and insufficient coverage during training.

Additionally, our current training methodology utilizes original hairstyle geometries without applying transformative augmentations such as squeezing, stretching, cutting, curling, or physical simulation. Exploration of more sophisticated data augmentation strategies would be valuable for future research, potentially enabling the generation of a wider variety of hairstyle configurations. Such augmentations could possibly enhance the diversity and versatility of the generated hairstyles, allowing the system to better accommodate the broad spectrum of stylistic preferences in character design contexts.
Regarding representation, while our approach provides a compact and competitive representation that effectively captures original data patterns within manageable token limits suitable for autoregressive generation, future work could investigate alternative representations such as generalized cylinders or explore higher compression ratios to further optimize the balance between representation fidelity and computational efficiency.

\section{More Results}
\xhdr{Choice of Discrete vs. Continuous Encoding and Decoding}
To validate the effectiveness of discrete encoding and decoding, we conducted an ablation study. For continuous encoding, we employed a two-layer MLP with a hidden size set to half of the output dimension and ReLU activation to convert floating-point attributes into embeddings. For continuous decoding, we switched to the MSE loss for training. As shown in Tab.~\ref{tab:encoding_decoding_ablation}, our proposed discrete encoding and decoding scheme both demonstrate advantageous performance.

\begin{table}[h]
\centering
\caption{Ablation study on encoding and decoding schemes.}
\resizebox{1.0\linewidth}{!}{
\begin{tabular}{l|l|cccc}
\toprule
\textbf{Encoding} & \textbf{Decoding} & CD~$\downarrow$ & EMD~$\downarrow$ & Hausdorff~$\downarrow$ & Voxel-IoU~$\uparrow$ \\ \midrule
Continuous & Continuous & 0.1079 & 0.0653 & 0.2748 & 0.2809 \\
Continuous & Discrete & 0.0296 & 0.0297 & 0.0980 & 0.5110 \\
Discrete & Continuous & 0.0203 & 0.0190 & 0.0850 & 0.5591 \\
Discrete & Discrete & \textbf{0.0117} & \textbf{0.0128} & \textbf{0.0497} & \textbf{0.7566} \\
\bottomrule
\end{tabular}
}
\label{tab:encoding_decoding_ablation}
\end{table}

\begin{figure*}[t]
\centering
\includegraphics[width=0.95\linewidth]{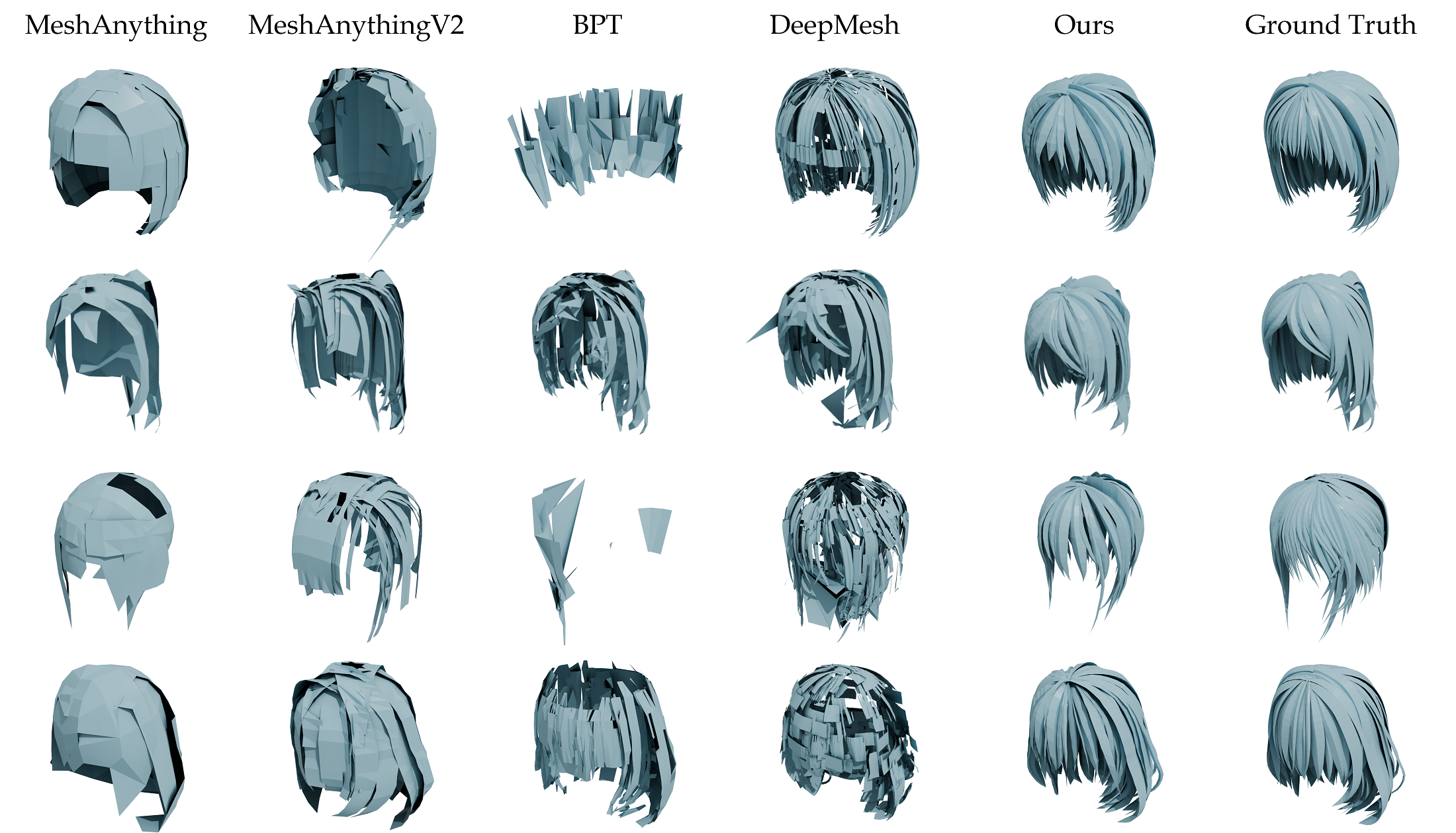}
\caption{More qualitative comparisons with other methods.}
\label{fig:main3}
\end{figure*}

\begin{figure*}[t]
\centering
\includegraphics[width=0.95\linewidth]{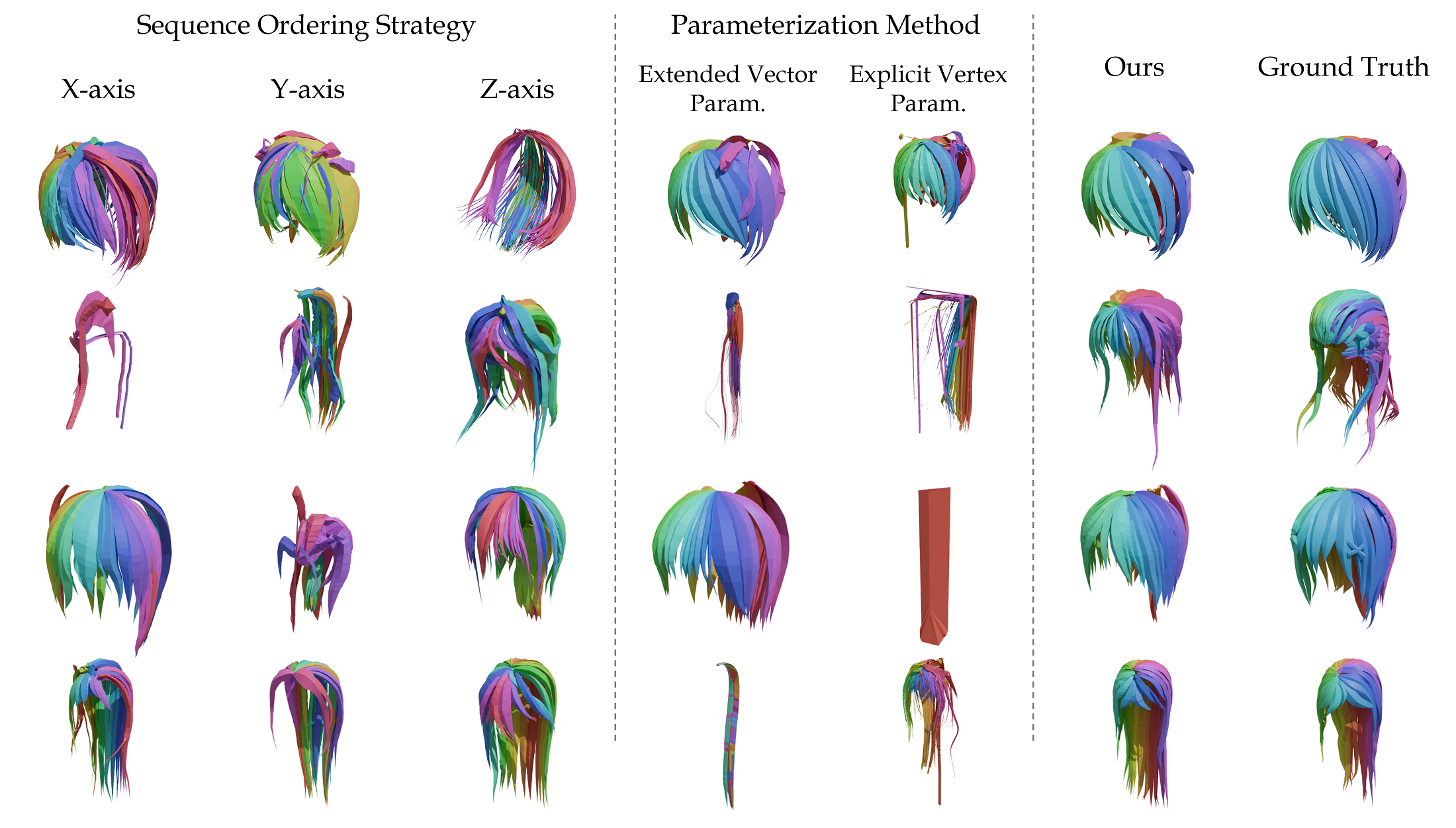}
\caption{More qualitative comparisons of ablation study. Colors indicate different hair strands.}
\label{fig:abl2}
\end{figure*}

\xhdr{More Qualitative Comparisons}
We provide more qualitative comparisons with other shape-conditioned 3D mesh generation methods in Fig.~\ref{fig:main3} and more qualitative comparisons of the ablation study in Fig.~\ref{fig:abl2}.

\end{document}